\documentclass[conference]{IEEEtran}
\IEEEoverridecommandlockouts
\usepackage{amsmath,amssymb,amsfonts}
\usepackage{graphicx}
\usepackage{textcomp}
\usepackage{xcolor}
\usepackage{booktabs} % For formal tables
\usepackage[linesnumbered,ruled,vlined]{algorithm2e}
\usepackage{booktabs}
\usepackage{amsmath}
\usepackage{graphicx}
\usepackage{subfigure}
\usepackage{cite}
\usepackage{multirow}
\usepackage{amsfonts,bm}
\usepackage{color}
\usepackage{braket}
\usepackage{mathtools}
\usepackage{enumerate}
\usepackage{enumitem}
\usepackage{empheq}
\usepackage{calc}
\usepackage{balance}
\usepackage{dsfont}
\usepackage{textcomp} 
\usepackage{color, colortbl}
\usepackage[first=0,last=9]{lcg}
\def\BibTeX{{\rm B\kern-.05em{\sc i\kern-.025em b}\kern-.08em
    T\kern-.1667em\lower.7ex\hbox{E}\kern-.125emX}}
\begin{document}

\title{Physics-Guided Recurrent Graph Model for Predicting Flow and Temperature in River Networks \\

% jzwart@usgs.gov,
% jsadler@usgs.gov,
% aappling@usgs.gov,
% soliver@usgs.gov,
% markstro@usgs.gov,
% willa099@umn.edu,
% xu000114@umn.edu,
% stei0062@umn.edu,
% jread@usgs.gov,
% kumar001@umn.edu

\author{Xiaowei Jia$^{1}$,  Jacob Zwart$^2$, Jeffrey Sadler$^2$, Alison Appling$^2$, Samantha Oliver$^2$, Steven Markstrom$^2$,\\ Jared Willard$^3$, Shaoming Xu$^3$, Michael Steinbach$^3$, Jordan Read$^2$, and Vipin Kumar$^3$\\
\small\baselineskip=9pt $^1$ University of Pittsburgh \\
\small\baselineskip=9pt $^2$U.S. Geological Survey\\
\small\baselineskip=9pt $^3$ University of Minnesota\\
\small  $^1$ xiaowei@pitt.edu, $^2$\{jzwart,jsadler,aappling,soliver,markstro,jread\}@usgs.gov, $^3$\{willa099,xu000114,stei0062,kumar001\}@umn.edu
}

%{\small \textsuperscript{*}Note: Sub-titles are not captured in Xplore and
}

% \and
% \and
% \and
% \and
% \and

\maketitle

\begin{abstract}
This paper proposes a physics-guided machine learning approach that combines advanced machine learning models and physics-based models to improve the prediction of water flow and temperature in river networks.  We first build a recurrent graph network model to capture the interactions among multiple segments in the river network. Then we present a pre-training technique which %leverages the simulation data generated by 
transfers knowledge from physics-based models to initialize the machine learning model and  learn the physics of streamflow and thermodynamics. We also propose a new loss function that balances the performance over different river segments. We demonstrate the effectiveness of the proposed method in predicting temperature and streamflow in a subset of the Delaware River Basin. In particular, we show that the proposed method brings a 33\%/14\% improvement over  the state-of-the-art physics-based model and 24\%/14\% over traditional machine learning models (e.g., Long-Short Term Memory Neural Network) in temperature/streamflow prediction using very sparse (0.1\%) observation data for training. %The proposed pre-training strategy also helps improve the prediction accuracy even with limited observation data. Moreover, the 
The proposed method has also been shown to produce better performance when generalized to different seasons or  river segments with different streamflow ranges.
\end{abstract}

% \begin{IEEEkeywords}
% Land Cover Detection, Domain Adaptation, Sequential Data
% \end{IEEEkeywords}

\section{Introduction}

Machine learning (ML) models, which have found immense success in commercial applications, e.g., computer vision and natural language processing, are beginning to play an important role in advancing scientific discovery~\cite{graham2008big_backup, jonathan2011special, sejnowski2014putting}. Given their power in automatically  learning from observation data, ML models are particularly promising in scientific problems involving complex processes that are not completely understood by our current body of knowledge.  %However, the notion of black-box application of data science has met with limited success in scientific domains (e.g., \cite{caldwell2014statistical,lazer2014parable, marcus2014eight}). 
However, scientific problems often involve non-stationary relationships among physical variables which can change over space and time. In the absence of adequate information about the physical mechanisms of real-world processes, traditional ML approaches are prone to false discoveries because it is difficult to capture these complex relationships solely from data.
%scientific problems often involve non-stationary dependencies among physical variables which may also change over space and time. Such relationships can be difficult to capture solely from observation data using standard ML models. 
Moreover, the data available for many scientific problems is far smaller than what is needed to effectively train advanced ML models.

The focus of this paper is on modeling physical systems that have multiple interacting processes. 
%In this paper, we propose an alternative ML-based solution to improve the predictions for complex systems with interacting components. 
In particular, we consider the application of predicting flow and temperature in river networks for both observed and unobserved river segments. %Prediction of flow and temperature at unobserved stream segments is critical for monitoring aquatic life and understanding biochemical cycling. This problem has been recognized as a major challenge in the scientific community~\cite{hrachowitz2013decade,sivapalan2003prediction}. 
In this problem, segments in the river network can show different flow and thermodynamic patterns driven by differences in catchment characteristics (e.g. slope, soil characteristics) %interacting with 
and meteorological drivers (e.g. temperature, precipitation). %with different input physical drivers. Different river segments 
These segments also interact with each other through the water advected from upstream to downstream segments. Moreover, there are often only a handful of river segments in a network that are monitored; thus there is limited data to train ML models.   
%while having observation data available only at certain river segments which are under frequent monitoring. This 
Accurate prediction of streamflow and water temperature aids in decision making for resource managers, establishes relationships between ecological outcomes and streamflow or water temperature, and helps predict other biogeochemical or ecological processes. For example, drinking water reservoir operators in the Delaware River Basin need to supply safe drinking water to New York City while also maintaining sufficient streamflow and cool water temperatures in the river network downstream of the reservoirs~\cite{ravindranath2016environmental}. Accurate prediction of streamflow and water temperature %with the river network 
helps managers optimize when and how much water to release downstream to maintain the flow and temperature regimes.

In scientific domains, physics-based models are often used to study engineering and environmental systems. Even though these models are based on known physical laws that govern relationships between input and output variables, most physics-based models are necessarily approximations of reality due to incomplete knowledge of certain processes or omission of processes to maintain computational efficiency. %These approximations introduce  inherent bias.  The limitations of physics-based models cut across discipline boundaries and are well known in the scientific community; e.g., see a series of debate papers in hydrology~\cite{lall2014debates, gupta2014debates, mcdonnell2014debates}. 
In particular,  existing physics-based approaches for predicting %both observed and unobserved river segments 
%river segments 
river networks simulate the internal distribution of target variables (e.g., streamflow and temperature) based on general physical relationships such as energy and mass conservation. However, %given that some processes are not fully understood and many physical variables cannot be easily measured, 
the model predictions still rely on qualitative parameterizations (approximations) based on soil and surficial geologic classification along with topography, land cover and climate input. Hence, such models %are  under-constrained and 
can only provide sub-optimal prediction performance. Furthermore, calibration of  physics-based models for river networks is often extremely time intensive due to interactions among parameters within segments and across segments and also requires expert knowledge of the system and model to calibrate successfully. The limitations of physics-based models cut across discipline boundaries and are well known in the scientific community; e.g., see a series of debate papers in hydrology~\cite{lall2014debates_backup, gupta2014debates_backup, mcdonnell2014debates_backup}.  %the parameters still interacted with each other for GLM but for PRMS-SNTemp there are interactions among parameters within segments and across segments making the calibration a bit slow \yell{Why is calibration for such complex systems extremely time consuming? } 

%Prior research has also investigated ways to leverage the knowledge encoded by physics-based models to guide the design and the training of ML models~\cite{jia2019physics,read2019process,karpatne2017theory}. These existing approaches aim to predict target variables for a single instance (e.g., a lake system) and do not explicitly model interaction among different processes. 

Intuitively, we can model each river segment independently by an individual ML model such as a Recurrent Neural Network (RNN). However, this approach has two major limitations: %in real-world settings:  %Interactions amongst river segments can be handled by explicitly incorporating the information from upstream ML models as inputs to downstream ML models. 
1) In a river network, there exist many different types of river segments with diverse catchment characteristics (e.g., slopes, elevation, etc.). Note that most of the segments are poorly observed or completely unobserved, which makes it impossible to build a purely data driven model for each segment separately. 2) The individual models may ignore the rich spatial and temporal contextual information, e.g., how the streamflows from upstream segments impact the water temperature in downstream segments in the next few days.

The first issue could be partly addressed by pre-training the ML model using simulation data produced by physics-based models,  
%(which may not good by Itemset but can still help initialize the ML model). B
but such pre-trained  ML models  still need some observations for refinement~\cite{jia2019physics_backup,read2019process_backup}. In particular, for unobserved river segments, the performance of pre-trained ML models can be no better than physics-based models that can have rather poor performance. %Hence, a global model with shared parameters for all the river segments would be preferred. Besides, modeling the spatial and temporal context (i.e., the second issue) is critical for the global ML model since it can enable learning of different behavior patterns even for two data points with identical input features but from different river segments and time steps. This 
Addressing the second issue will require the development of sophisticated ML architecture that can leverage latent information that is transferred  across river segments. %However, the latent information extracted by ML models using only sparse observation data can be less representative and also inconsistent with the physical knowledge. This can be improved by using auxiliary data sources with physical knowledge. 

%However, this is not feasible  given that only a few  river segments are monitored and thus we have no observation data for training individual models for the remaining river segments. Although we can possibly train individual ML models using simulation data generated by physics-based models, they tend to perform badly since  physics-based models  cannot be calibrated for  unobserved segments.  A commonly used approach is to build a global model for the entire river network and the model parameters are shared across different river segments. Given the heterogeneity across the river network, it is critical for the global to capture the spatial and temporal contextual information that enables learning of different behavior patterns even for two identical data points from different river segments and time steps.

% , and the interaction is captured via graph neural network.  Discuss advantages/disadvantages of these and motivate why you are focusing only on approach 2 in this paper.   After this you can discuss challenges such as those discussed 

% For links with observations, SDM/WRR approach can be used.  For links with no observations, models can only be pre trained using  SNtemp. Interaction is handled by explicit input and out of flow across links. 

%To improve predictions over the entire river networks, 
In this paper, we propose a global model, Physics Guided Recurrent Graph Networks (PGRGrN), which is applied to  all the river segments. %, which uses the time series of observation data along with a physics-based model providing the simulated internal distributions of physical variables to serve as an auxiliary data source. We introduce a new 
The architecture of PGRGrN is based on  Recurrent Neural Networks (RNN) and Graph Convolutional Neural Networks (GCN), which explicitly captures the spatial interactions among different river segments as well as their  temporal dynamics. Modeling of the spatial and temporal context is critical for the global ML model as it enables learning of different behavior patterns for different river segments even when they have similar input features on certain dates.

%even for two data points with identical input features but from different river segments and time steps.

%Here the real challenge is whether this model can truly extract physical variables that are transferred amongst different river segments and thus improve on the ability to adequately capture how segment-to-segment interactions  influence the change of target variables over the entire watershed network. 
Design of such an architecture for this application faces two challenges. First, 
%While GCN-based models has shown effective in many commercial applications (e.g., social network, recommender systems), they 
existing GCN-based models extract abstract latent variables (i.e., graph embeddings) to propagate over the networks but do not explicitly incorporate the prior physical knowledge about the interactions among different nodes. Such latent variables can become less informative when they are learned from sparser and less representative observation data, which can make the GCN model not generalizable. To address this challenge, we propose to utilize the intermediate variables simulated by the physics-based model to guide the learning process of the graph neural networks. This approach aims to %ensure that the output of each network component (i.e., the modeling component of each river segment) has 
enforce a physical interpretation to latent variables learned from each river segment  by transferring the prior knowledge encoded by the physics-based model to the proposed ML model. %In particular, we add supervision to the latent variables of the ML model using intermediate physical variables simulated by physics-based models. %to pre-train the ML model by adding supervision to the latent variables that are transferred across different river segments. 
Our experimental evaluation shows that this strategy is effective in initializing the ML model, which can then be fine-tuned using observations from the river network. %Once the ML model is pre-trained using a large amount of simulation data, we then fine-tune the model to its target solution using observations from the river network.  

The second challenge is that target variables can vary drastically across different processes of a complex system. For example, streamflow can vary by several orders of magnitude across segments in a river network. % while temperature may vary only a few degrees. % streamflow is very sensitive to a slight difference in geometric structure of river segment (e.g., depth and width). 
When we train a global ML model over the entire river network, the training process using traditional loss functions (e.g., mean squared loss) tends to optimize the overall performance over training data available for the entire river networks, %. %. The training process using traditional loss functions 
%The training process using traditional loss functions is very likely to be 
and thus can be dominated by river segments that contribute most to the overall error (e.g., segments with high streamflow).  %Popular choices of loss functions include cross-entropy (classification) and RMSE (regression). Although the optimization using the general loss function usually results in a (local) minimum in the training set, it may not produce desired results for sub-populations from a physical perspective. 
However, it is also important to accurately predict river segments with lower streamflow, as accurate prediction for these segments provides important information regarding the  habitat for aquatic life and aquatic biogeochemical cycling. To address this challenge, we design a new loss function to ensure that the global ML model can simultaneously capture the local patterns of all the different segments. The local patterns of each segment can be extracted using an individual ML model trained only for this segment using  simulation data (which is plentiful).  
%\yell{define global?}
%One intuitive way to overcome this problem is to train individual ML models for each river segment. Although this is not feasible given that observation data are only available for certain segments, w
%Specifically, we extract local patterns of each individual segment using the simulation data generated by physics-based models for every segment and date.  %phyics-based models to generate simulation data at every segment and date to . 
Then during the training of the global ML model, we use a distance-based loss function, the contrastive loss function,  to enforce its consistency with the extracted local patterns. %This loss function can be widely applied to scientific problems where target variables vary a lot across different processes. % multiple physical processes produce target variables at different scales. % for the training of the  global ML model. %to conform to the extracted local patterns. 

We implement our proposed method in a real-world dataset collected over 36 years from the Delaware River Basin located in the Northeast United States and demonstrate our method's superior predictive performance over existing methods. Moreover, we show that the proposed method produces much better prediction performance using sparse observations and also has better generalizability.

Our contributions can be summarized as follows:
\begin{itemize}
    \item We introduce a new recurrent graph network architecture to model a river network with interacting river segments.
    \item We leverage knowledge from physics-based models to guide ML models for extracting latent variables, %via a new pre-training strategy. %that uses simulation data to initialize 
    which helps initialize the model while enforcing consistency with known physical relationships amongst different processes.
    \item We propose a new contrastive loss function that balances the prediction performance over different river segments.
    \item We evaluate the utility in the context of an ecologically and societally relevant problem of monitoring river networks.
\end{itemize}

\section{Related Work}
% pure ML model on physics and PGML
%ML models have shown a lot of promise in modeling environmental systems~\cite{ham2019deep,karpatne2017theory}. Several r
Recent works have shown the promise of integrating physics into ML models in improving the predictive performance and generalizability in scientific problems.   %~\cite{read2019process}. %In particular, previous research has discussed several ways of integrating physics into ML models, 
This is commonly conducted in several ways, including %such as 
developing new model architectures~\cite{anderson2019cormorant_backup,muralidhar2018incorporating_backup}, applying additional loss functions~\cite{jia2019physics_backup,read2019process_backup}, and modeling prediction residuals~\cite{wang2017physics_backup,liu2019multi}. 
% PGML-architecture 
When applied to systems with interacting processes, %it is critical to ensure that 
ML models are expected to have sufficient capacity %so that they can 
to model such interactions.  %the interactions amongst different processes. 
New ML architectures have been designed to enforce known physical relationships %in defining the %relationships 
%from input to the output
among multiple internal processes that jointly convert inputs to outputs~\cite{muralidhar2018incorporating_backup,anderson2019cormorant_backup,park2019physics}, thus reducing the space for searching parameters. 
ML models have also shown great potential in modeling river networks~\cite{kratzert2019toward_backup,moshe2020hydronets_backup}. For example, Moshe et al.~\cite{moshe2020hydronets_backup} propose HydroNets,  which combines the information from each river segment and its upstream segments for improving streamflow predictions. It also learns local patterns for each basin by introducing basin-specific model layers in addition to the global model. This method focuses on predicting basins that are well monitored and it remains limited in generalizing to different scenarios or learning with less data. In contrast, we leverage the prior physical knowledge to learn latent variables that make the model more generalizable. % when   using prior physical knowledge, this model can still have limited generalizability when learned from less representative training data.  

% Graph NN
The Graph Convolutional Networks (GCN) model %has been widely used in commercial applications and it 
has proven to be effective in automatically extracting latent factors that influence the neighbors in a graph. The use of GCN has also shown improved prediction accuracy in several scientific problems~\cite{qi2019hybrid,xie2018crystal,zhu2020understanding}. However, the information propagated amongst nodes in GCN is essentially an  abstract representation learned by end-to-end training. 
Such abstract representations are not meant to enforce consistency with known physical relationships among different processes, such as in river networks. 

%Such abstract representation may not contain sufficient physical knowledge to be consistent with known physical relationships between different components.

% simulation data
Simulation data have been used to assist in training ML models. Since many ML models require an initial choice of model parameters before training, researchers have explored different ways to physically inform a model starting state. Poor initialization can cause models to anchor in local minima, which is especially true for deep neural networks. %\textit{Transfer learning} can effectively tackle this issue, where the pre-trained models from a related task are fine-tuned with limited training data to fit the desired task. 
One way to harness physics-based modeling knowledge is to use the physics-based model's simulated data to pre-train the ML model, which also alleviates data paucity issues. Jia \textit{et al.} extensively discuss this strategy~\cite{jia2019physics_backup}. They pre-train their Physics-Guided Recurrent Neural Network (PGRNN) models for lake temperature modeling on simulated data generated from a physics-based model and fine-tune it with little observed data. They show that pre-training %, even using data from a physical model with an uncalibrated set of parameters, 
can %still 
significantly reduce the training data needed for a quality model. In addition, Read et al.~\cite{read2019process} show that such models are able to generalize better to unseen scenarios than pure physics-based models. %In particular, they show that an ML model pre-trained with data from a poorly calibrated physical simulation and refined using observations from the spring, fall, and winter seasons was able to greatly outperform the physical model calibrated using the same set of observations when predicting during summer seasons (for which, observations were not made available to either approach).

% Besides direct incorporation of physical laws as additional constraints, it is also possible to indirectly incorporate physical knowledge encoded in physics-based/mechanical models. These models are built based on known physical laws and thus contain rich information about physical systems. The incorporation of the knowledge encoded by physics-based models not only enables learning physically-consistent solutions, but also allows better discovery of inherent physical processes. Although these physics-based models often include inaccurate or overly-complex processes due to incomplete knowledge of physics or unknown practical factors, they can be used to generate large amounts of synthetic ground truth training data to pre-train the ML model. The idea here is that training from synthetic data generated by imperfect physical models may allow the ML model to get close enough to the target solution, so only a small amount of observed data (ground truth labels) is needed to further refine the model.

\section{Problem Definition and Preliminaries}
\subsection{Problem definition}
Our objective is to model the dynamics of temperature and streamflow in a set of connected river segments that together form a river network.  %In this problem, we are provided with data from multiple segments in the river networks. 
The connections amongst these river segments can be represented in a graph structure $\mathcal{G} = \{\mathcal{V},\mathcal{E},\textbf{A}\}$, where $\mathcal{V}$ represents the set of river segments and $\mathcal{E}$ represents the set of connections amongst river segments. Specifically,  we create an edge $(i,j)\in\mathcal{E}$ if the segment $i$ is anywhere upstream of the segment $j$. %We include %It is worth mentioning %that we also include the segments that are upstream of the segment $j$ but 
%the segment  may not necessarily be the  direct  upstream segment to the segment $j$ %in the edge set $\mathcal{E}$ 
%because these segments can still have impact to the downstream segment $j$. 
The matrix $\textbf{A}$ represents the adjacency level between each pair of segments, i.e.,  $\textbf{A}_{ij}=0$ means there is no edge from the segment $i$ to the segment $j$ and a higher value of $\textbf{A}_{ij}$ indicates that the segment $i$ is closer to the segment $j$ in terms of the river distance. 
More details of how we generate the adjacency matrix are discussed in Section~\ref{sec:dataset}. 

For each river segment $i$, we have access to its input features at multiple time steps $\textbf{X}_i=\{\textbf{x}_{i}^{1}, \textbf{x}_{i}^2, ..., \textbf{x}_{i}^T\}$.  The input features $\textbf{x}_{i}^t$ are a $D$-dimensional vector, which include meteorological drivers, geometric parameters of the segments, etc. (more details can be found in Section~\ref{sec:dataset}). We also have a set of observed target variables $\textbf{Y}=\{y^t_i\}$ but they are only available   for  certain time steps $t\in \{1,...,T\}$ and certain segments $i\in\{1,...,N\}$.

%Although we have access to limited observation data, we can run physics-based models to simulate target variables $\{\tilde{y}_i^t\}$ at every time step and for every segment. It is noteworthy that the simulation data are imperfect due to the inherent bias of the uncalibrated physics-based model, but they still follow many general physical relationships that are used to build the physics-based model.

% \yell{Introduction of simulation data here.}

\subsection{Physics-based Streamflow and Temperature Model}
% \yell{Jake: please help check this and add more details if necessary. }

The Precipitation-Runoff Modeling System (PRMS)~\cite{markstrom2015prms} and the coupled Stream Network Temperature Model (SNTemp)~\cite{sanders2017documentation_backup} is a physics-based model that simulates daily streamflow and water temperature for river networks, and other variables. PRMS is a one-dimensional, distributed-parameter modeling system that translates spatially-explicit meteorological information into water information including evaporation, transpiration, runoff, infiltration, groundwater flow, and streamflow. PRMS has been used to simulate catchment hydrologic variables at regional~\cite{lafontaine2013application} to national scales~\cite{regan2018description_backup} in support of resource management decisions, among other applications. The SNTemp module for PRMS %was based on the standalone, physics-based stream temperature model developed by Theurer et al.~\cite{theurer1984instream}, and 
simulates mean daily stream water temperature for each river segment by solving an energy mass balance model which accounts for the effect of inflows (upstream, groundwater, surface runoff), outflows, and surface heating and cooling on heat transfer in each stream segment. The SNTemp module is driven by the same meteorological drivers used in PRMS and also driven by the hydrologic information simulated by PRMS (e.g. streamflow, groundwater flow). %For our application, we use a cutout of the National Hydrologic Model application of PRMS for the Delaware River Basin (our modeling domain), which includes a national-scale geospatial modeling fabric and spatially-explicit parameterization of various PRMS parameters~\cite{regan2018description}. 
Calibration of PRMS-SNTemp is extremely time-consuming because it involves a large number of parameters (84 parameters) and the parameters interact with each other both within segments and across segments.

%It also includes a precipitation-runoff modeling system which is developed to evaluate the response of various combinations of climate and land use on streamflow and general watershed hydrology based on known physical processes.  % and assumes that all input data, including meteorological and hydrological variables, can be represented by 24-hour averages.

%is a mechanistic, one-dimensional heat transport model that predicts the daily mean and maximum water temperatures as a function of stream distance and environmental heat flux. Net heat flux is calculated as the sum of heat to or from long-wave atmospheric radiation, direct short-wave solar radiation, convection, conduction, evaporation, streamside vegetation (shading), streambed fluid friction, and the water's back radiation . The heat flux model includes the incorporation of groundwater influx. The heat transport model is based on the dynamic temperature-steady flow equation and assumes that all input data, including meteorological and hydrological variables, can be represented by 24-hour averages.

\subsection{Recurrent Neural Networks and Long-Short Term Memory}
The RNN model has been widely used to model the temporal patterns in sequential data. The RNN model defines  transition relationships for the extracted hidden representation through a recurrent cell structure. In this work, we adopt the Long-Short Term Memory (LSTM) cell which has proven to be effective in capturing long-term dependencies. %In essence, the Long-Short Term Memory (LSTM) model defines a transition relationship for hidden representation $h_{{t}}$ %at each time step 
%through an LSTM cell, %. This LSTM cell 
%which 
The LSTM cell combines the input features $\textbf{x}^{t}$ at each time step and the inherited information from previous time steps. Here we omit the subscript $i$ as we do not target a specific river segment.

% \begin{figure} [!h]
% \centering
% %\raggedleft
% \label{fig:b}{}
% \includegraphics[width=0.75\columnwidth]{lstm_mod.png}
% \caption{The structure of LSTM cell.}
% \label{lstm_cell}
% \end{figure}

Each LSTM cell has a cell state $\textbf{c}^t$, which serves as a memory and allows %the hidden units $h^t$ to reserve
preserving information from the past. 
Specifically, the LSTM first generates a candidate cell state $\bar{\textbf{c}}^t$ by combining $\textbf{\textbf{x}}^t$ and the hidden representation at previous time step $\textbf{\textbf{h}}^{t-1}$, as follows: %into a $\text{tanh}(\cdot)$ function:
\begin{equation}
\small
\begin{aligned}
\bar{\textbf{c}}^t &= \text{tanh}(\textbf{W}_c^h \textbf{h}^{t-1} + \textbf{W}_c^x \textbf{x}^t+\textbf{b}_c).
\end{aligned}
\end{equation}

% Hereinafter we omit the bias terms as they can be included in the weight matrices.

Then the LSTM generates a forget gate $f^t$, an input gate  $g^t$, and an output gate  via sigmoid function $\sigma(\cdot)$, as follows:
%\vspace{-.07in}
\begin{equation}
\small
\begin{aligned}
%f^t = \sigma(W^f_h h^{t-1} + W^f_x x^t).
\textbf{f}^t &= \sigma(\textbf{W}_f^h \textbf{h}^{t-1} + \textbf{W}_f^x \textbf{x}^t+\textbf{b}_f),\\
\textbf{g}^t &= \sigma(\textbf{W}_g^h \textbf{h}^{t-1} + \textbf{W}_g^x \textbf{x}^t+\textbf{b}_g),\\
\textbf{o}^t &= \sigma(\textbf{W}_o^h \textbf{h}^{t-1} + \textbf{W}_o^x \textbf{x}^t+\textbf{b}_o).
\end{aligned}
\end{equation}

The forget gate  is used to filter the information inherited from $\textbf{c}^{t-1}$, and the input gate  is used to filter the candidate cell state at $t$. Then we compute the new cell state and the hidden representation as follows: %and the hidden representation as: 
\begin{equation}
\small
\begin{aligned}
\textbf{c}^t &= \textbf{f}^t\otimes \textbf{c}^{t-1}+\textbf{g}^t\otimes\bar{\textbf{c}}^t,\\
\textbf{h}^t &= \textbf{o}^t\otimes \text{tanh}(\textbf{c}^t),
\end{aligned}
\label{hid}
\end{equation}
where $\otimes$ denotes the entry-wise product.
%\yell{change the notation to be subscript}

%Finally, the LSTM cell outputs the hidden representation as:
% \begin{equation}
% \small
%     \textbf{h}^t = \textbf{o}^t\otimes \text{tanh}(\textbf{c}^t).
%     \label{hid}
% \end{equation}

According to the above equations, we can observe that the computation of $\textbf{h}^t$ %not only depends on $\textbf{x}^{t}$, but also utilizes the inherited information from 
combines the information at current time step ($\textbf{x}^{t}$) and previous time step ($\textbf{h}^{t-1}$ and $\textbf{c}^{t-1}$), %. Hence, the hidden representation 
and thus encodes the temporal patterns learned from data. 

%In following sections, we represent this process as $h^{t} = \text{LSTM} (x^t| \mathcal{I}^{t-1})$, where $\mathcal{I}^{t-1}$ represents the inherited information from $t-1$.

% As we wish to conduct regression for continuous values, we generate the predicted temperature $\hat{y^t}$ at each time step $t$ via a linear combination of hidden units, as:
% \begin{equation}
% \small
%     \hat{\textbf{y}}^t = \textbf{W}_y \textbf{h}^t.
% \end{equation}
% %where $W_y$ is the weight matrix to transform hidden representation.

% We apply the LSTM model for each depth separately. Given the true observation $y_{d,t}$ at every time step and at every depth, our training loss is defined as:
% \begin{equation}
% \small
%     \mathcal{L}_{\text{RNN}} = \frac{1}{N T}\sum_i \sum_{t} (y_{i,t}-\hat{y}_{i,t})^2,
% \end{equation}
% where $N_d$ is the total number of different depths, and $T$ is the number of time steps.

\section{Method}
In this section, we describe the details of the PGRGrN method. We start with introducing the model architecture. Then we  discuss a strategy to help enforce physical relationships by leveraging the physical knowledge embedded in physics-based models. Finally, we introduce a contrastive loss function that attempts to ensure that  the model performance on individual river segments is not compromised while optimizing the performance on the entire set of segments.

\begin{figure} [!h]
\centering
\subfigure[]{ \label{fig:b}{}
\includegraphics[width=0.65\columnwidth]{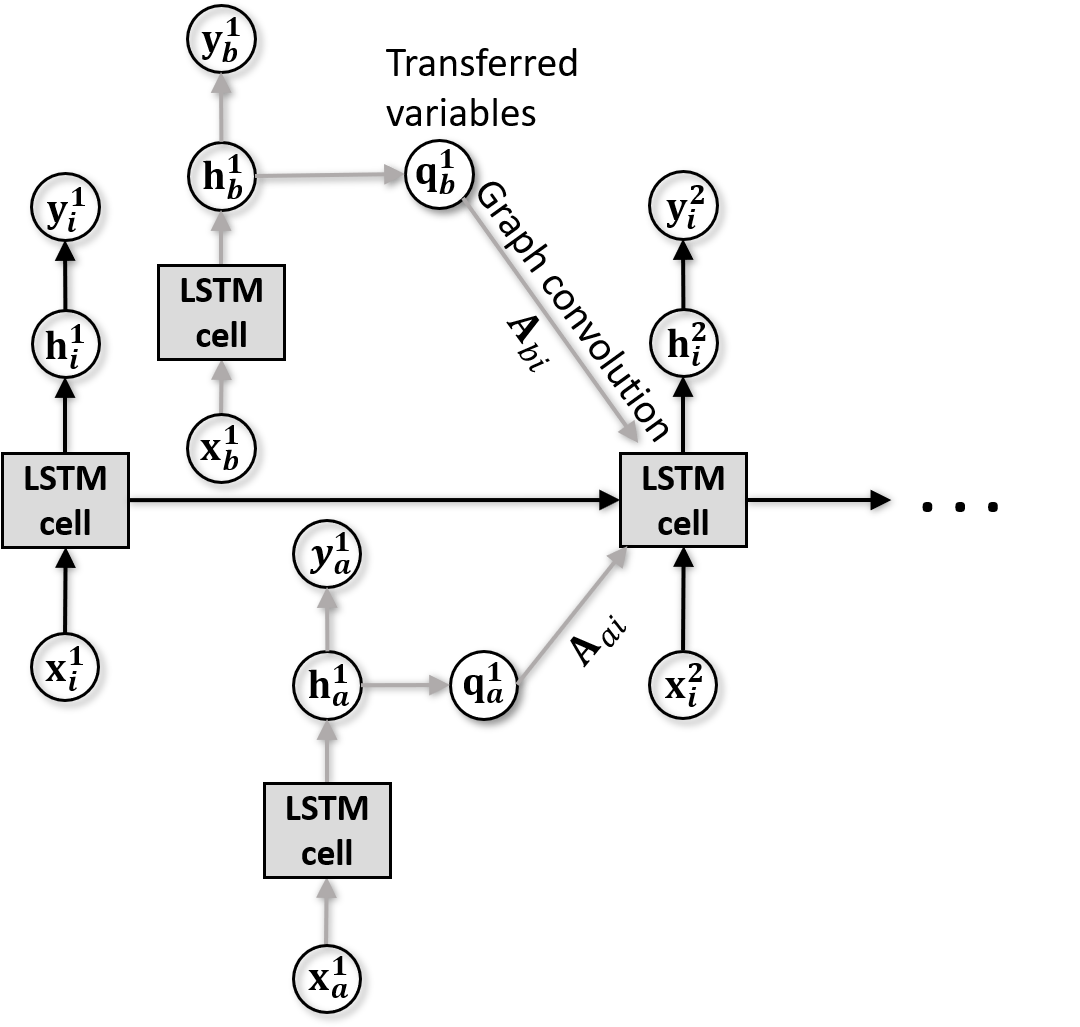}
}%\hspace{-.1in}
\subfigure[]{ \label{fig:b}{}
\includegraphics[width=0.28\columnwidth]{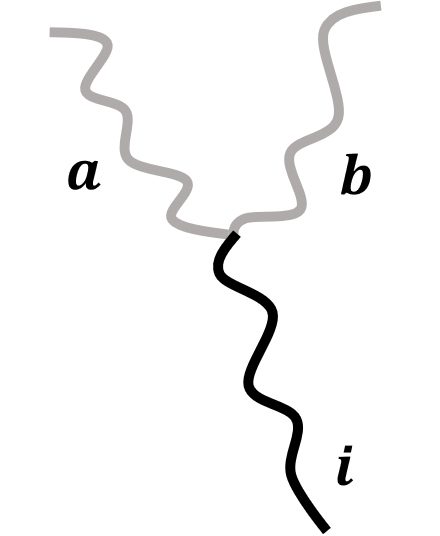}
}
\caption{(a) The RGrN architecture for an example river network with three segments (b). The segment $a$ and the segment $b$ are upstream of the segment~$i$. The grey arrows indicate the modeling components for upstream segments.}
\label{flow_chart}
\end{figure}

\subsection{Recurrent Graph Network}
In a river network, most river segments are poorly observed or completely unobserved. To this end, we introduce a global ML model architecture, Recurrent Graph Network (RGrN), which is trained using data collected from all the river segments. %The global ML model has its model parameters shared across different segments. Our proposed PGRGrN model is a global ML model since we apply the trained model to predict streamflow and temperature for the entire river networks using the same set of model parameters.  
%PGRGrN performs better than RNN because it utilizes upstream-downstream dependencies which are critical for accurate accounting of temperature or streamflow. 
Effective modeling of river segments requires the ability to capture their temporal dynamics and the influence received from upstream segments. Hence, we %build a new recurrent graph model that  incorporates 
incorporate the information  from both previous time steps and neighbors (i.e., upstream segments) when modeling each segment (Fig.~\ref{flow_chart}). 

Here we describe the recurrent process of generating the hidden representation $\textbf{h}^t$ from $\textbf{h}^{t-1}$, and we repeat this process for the entire sequence from $t=2$ to $T$ ($\textbf{h}^1$ learned from an LSTM model). For each river segment $i$ at time $t-1$, the model extracts latent variables %from this segment 
which contain relevant information to pass to its downstream segments. We refer to these latent variables as transferred variables. For example, the amount of water advected from this segment and its water temperature can directly impact the change of water temperature for its downstream segments.  %.   
%Formally, we generate transferred variables 
 %that contain the relevant information to pass to its downstream segments. The transferred variables are computed
%and they are generated 
We generate the transferred variables $\textbf{q}_{i}^{t-1}$ from the hidden representation $\textbf{h}_{i}^{t-1}$ , as follows: %follows:
\begin{equation}
\small
    \textbf{q}_{i}^{t-1} = \text{tanh}(\textbf{W}_q\textbf{h}_{i}^{t-1}+\textbf{b}_q),
\end{equation}
where $\textbf{W}_q$ and $\textbf{b}_q$ are model parameters that are used to  convert the hidden representation to transferred variables. 

After gathering the transferred variables for all the segments, we develop a new recurrent cell structure for each segment $i$ that integrates the transferred variables from its upstream segments into the computation of the cell state $\textbf{c}_i^t$. This can be expressed as follows: 
\begin{equation}
\small
    \textbf{c}_{i}^{t} = \textbf{f}_{i}^{t}\otimes (\textbf{c}_{i}^{t-1}+\sum_{(j,i)\in \mathcal{E}}\textbf{A}_{ji}\textbf{q}_{j}^{t-1})+\textbf{g}_{i}^t\otimes\bar{\textbf{c}}_{i}^t
    \label{conv}
\end{equation}

%\yell{Mention the special case of head water. }
We can observe that the forget gate not only filters the previous information from the  segment $i$ itself but also from its neighbors (i.e., upstream segments). Each upstream segment $j$ is weighted by the adjacency level $\textbf{A}_{ji}$ between $j$ and $i$. When a river segment has no upstream segments (i.e., head water), the computation of %cell state in Eq.~\ref{conv} 
$\textbf{c}_i^t$ is the same as with the standard LSTM. In Eq.~\ref{conv}, we use $\textbf{q}_{j}^{t-1}$ %in Eq.~\ref{conv} 
%computed at 
from the previous time step %because there is usually 
because of the time delay in transferring the influence from upstream to downstream segments (the maximum travel time is approximately one day according to PRMS). %as PRMS segment lengths were generated to have a maximum travel time of approximately one day. 
We also discuss the %potential 
impact of increasing the time delay in Section~\ref{sec:sens}.

After obtaining the cell state, we can compute the hidden representation $\textbf{h}_i^t$ by following Eq.~\ref{hid}. Finally, we generate the predicted output from the hidden representation as follows:
\begin{equation}
\small
    \hat{\textbf{y}}_i^t = \textbf{W}_y \textbf{h}_i^t+\textbf{b}_y,
    \label{prd}
\end{equation}
where $\textbf{W}_y$ and $\textbf{b}_y$ are model parameters.

%We apply the proposed model to a network of river segments. 
After applying this recurrent process to all the time steps, we define a loss using true observations %Given the set of true observations 
$\textbf{Y}=\{\textbf{y}_{i}^{t}\}$ that are available at certain time steps and certain segments, as follows: % our training loss is defined as:
\begin{equation}
\small
    \mathcal{L}_{\text{RGrN}} = \frac{1}{|\textbf{Y}|} \sum_{\{(i,t)|\textbf{y}_{i}^{t}\in \textbf{Y}\}} (\textbf{y}_{i}^{t}-\hat{\textbf{y}}_{i}^{t})^2.
    \label{loss_PGRGrN}
\end{equation}

%The training of RGrN can be done using the standard back-propagation algorithm. 

\subsection{Transferring knowledge from physics-based models}
\label{sec:ptr}
The RGrN architecture has the capacity to model the latent information that is transferred across river segments. %so that downstream segments can receive the %relevant 
%information needed to estimate target variables. 
However,  training  RGrN directly in an end-to-end fashion can only learn an abstract representation for  transferred variables while ignoring their %potential 
physical interpretation. The transferred variables can become less informative when they are learned from sparser and less representative observation data. To this end, we introduce a new strategy to %initialize the proposed ML model by  enforcing 
enforce the prior physical relationships amongst different river segments which are encoded by physics-based models. %propose a pre-training strategy %with the aim to learn more  
%This strategy enhance the learning of transferred variables such that they are consistent with prior physical knowledge encoded by physics-based models. %(e.g. groundwater flow, subsurface flow) 
It helps make  RGrN model more generalizable and also reduces the amount of observation data required to train a high-quality model. This strategy can be applied to a wide range of scientific problems that are modeled as a set of interacting processes. %with interacting processes given access to the corresponding physics-based model to provide its encoded physical relationships and  simulation data.
%Our goal is to ensure that the transferred variables at each segment in our proposed ML model contains sufficient information to represent the intermediate physical variables so that the downstream segment can gather all the relevant information for capturing $F_{up}$.

We use the river temperature modeling as an example in this section to better illustrate the proposed strategy. %However, the proposed pre-training approach has a wide applicability to many scientific applications.
For each river segment in the network, the temperature change is driven by energy exchanges caused by solar radiation, rainfall, evaporation, conductive and convective heat transfer, and net heat advected into river segments (e.g., groundwater flow, upstream flow, downstream flow). These energy fluxes can be summarized into three categories %: incoming energy fluxes from solar radiation, rainfall and other natural sources ($F_{in}$), outgoing energy fluxes ($F_{OUT}$) and net heat advected  . We can summarize this 
and the process of temperature change conforms to  the equation as follows:
\begin{equation}
\small
    \Delta \text{Temperature} \propto F_{in}-F_{out}+F_{up},
\end{equation}
where $F_{in}$ denotes the incoming energy fluxes from solar radiation, rainfall and other natural sources, $F_{out}$ denotes the outgoing energy fluxes including long-wave emission, evaporation, conductive and convective heat transfer, and $F_{up}$ denotes the net heat advected into the river segment from upstream segments. The term $F_{up}$ can be estimated by a set of intermediate physical variables from upstream segments, which include upstream flow, upstream water temperature, relative humidity, and other physical characteristics. These intermediate physical variables can be simulated by PRMS-SNTemp internally. % and include upstream flow, upstream water temperature, relative humidity, and other physical characteristics.  %We call these variables as intermediate physical variables. 

Our goal is to ensure that the transferred variables $\textbf{q}_i^t$ at each segment %in our proposed ML model 
contain sufficient information to represent these intermediate physical variables so that  downstream segments can gather all the information needed for capturing $F_{up}$. For each segment $i$ at time $t$, we first simulate the set of intermediate variables $\textbf{s}_{i}^{t}$ by running %the 
PRMS-SNTemp. %model. 
Then we  %pre-train the ML model to 
use $\textbf{s}_i^t$ to add supervision on the transferred variables $\textbf{q}_{i}^{t}$ %to ensure 
such that we can extract $\textbf{s}_i^t$ from $\textbf{q}_{i}^{t}$. % contains sufficient information to represent the intermediate physical variables. %extract the intermediate physical variables from transferred variables $\textbf{q}_{i,t}$. 
More formally, we define a loss function on transferred variables as follows:
\begin{equation}
\small
    \mathcal{L}_{\text{trans}} = \frac{1}{N T} \sum_{i} \sum_t ||\textbf{s}_{i}^{t}-(\textbf{W}_s\textbf{q}_{i}^{t}+\textbf{b}_s)||^2,
    \label{ltrans}
\end{equation}
where $\textbf{W}_s$ and $\textbf{b}_s$ are model parameters that  convert transferred variables to the intermediate physical variables. We call this model as Physics-Guided Recurrent Graph Networks (PGRGrN) because we leverage the physical relationships between different river segments. 

Since the computation of loss $\mathcal{L}_\text{trans}$ does not require observation data, we can use it  to pre-train RGrN to enforce physical relationships. 
This pre-training method not only explicitly enforces the physical relationships among river segments, but also enables full usage of physical intermediates obtained from physics-based models to enhance the representation learned for $\textbf{q}_i^t$ and its previous layer $\textbf{h}_i^t$.  In particular, the intermediate physical variables used in this work include streamflow, stream temperature, relative humidity, cloud cover, groundwater and shallow subsurface flow, and surface runoff.

It is noteworthy that the model extracts the intermediate variables from transferred variables $\textbf{q}_i^t$ rather than forcing the transferred variables to be exactly the intermediate physical variables. Also, we optimize the loss $\mathcal{L}_\text{trans}$ in pre-training rather than using it as a regularizer in supervised learning.  %This not only ensures that $p_{i,t}$ contains sufficient information to enforce the known physical relationship, but 
In other words, the ML model is guided but not constrained by the physics-based model output, which allows for more flexibility to automatically learn information in $\textbf{q}_{i}^{t}$ that is poorly known or not yet discovered while also remaining helpful for modeling the interactions among river segments.

% Although we have access to limited observation data, we can run physics-based models (PRMS-SNTemp) to simulate target variables $\{\tilde{y}_i^t\}$ at every time step and for every segment. Despite the inherent bias of physics-based models, they still follow many general physical relationships that are used to build the physics-based model.

At the same time, we can run PRMS-SNTemp to simulate the final target variables corresponding to $\textbf{y}_{i}^{t}$. Here we use $\tilde{\textbf{y}}_{i}^{t}$ to represent the simulated target variables by PRMS-SNTemp. Although the simulated data are not  accurate reflection of the observation data, we can generate adequate simulations on every day and for every segment. The simulation data also follow many general physical relationships used to build the physics-based model. Given that observation data is often scarce, we can use simulated target variables %can help 
to initialize the model via pre-training. Hence, we define another pre-training loss  on target variables as follows:
\begin{equation}
\small
    \mathcal{L}_{\text{tar}} = \frac{1}{N T} \sum_{i} \sum_t (\tilde{\textbf{y}}_{i}^{t}-\hat{\textbf{y}}_{i}^{t})^2,
    \label{ltar}
\end{equation}

\begin{figure} [!t]
\centering
\includegraphics[width=0.7\columnwidth]{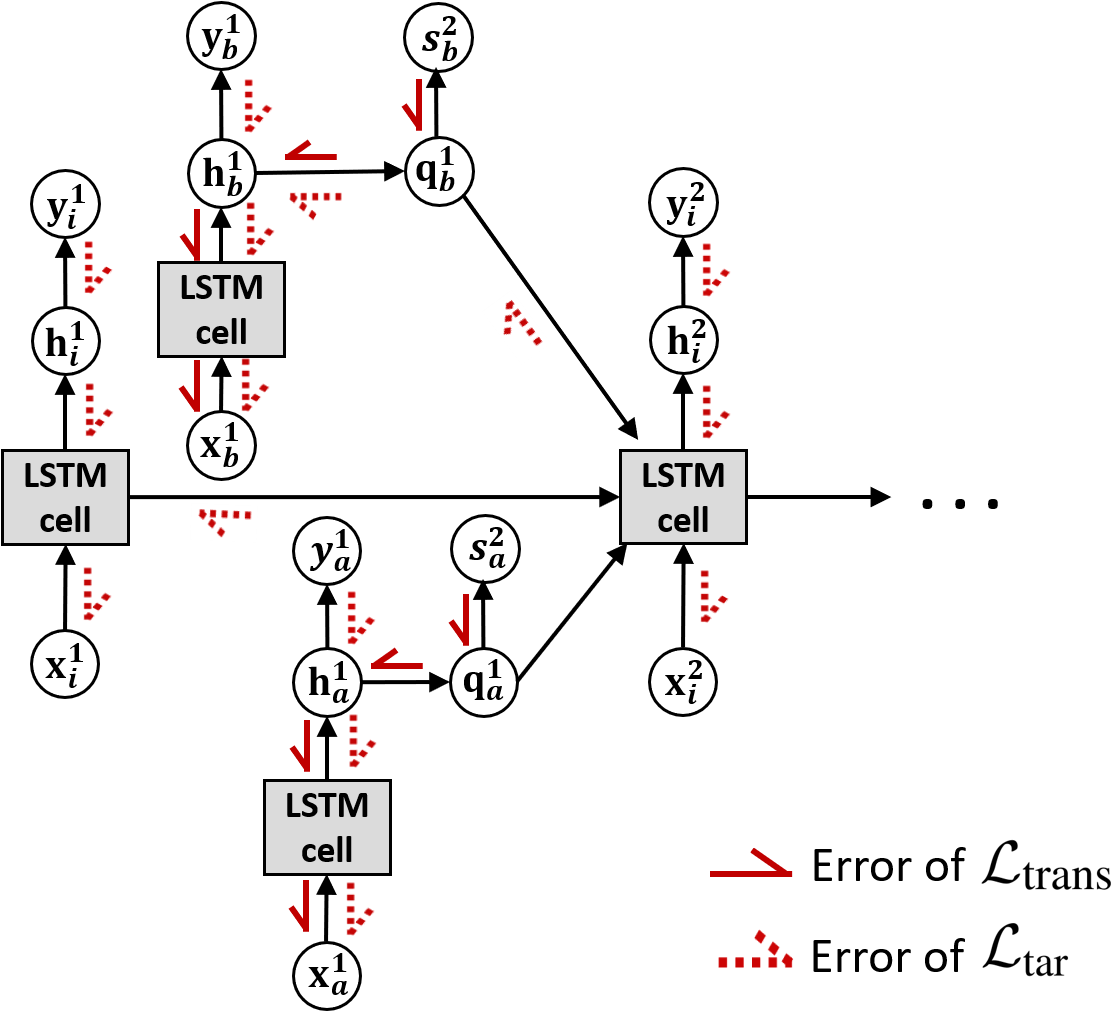}
\caption{Error back-propagation during pre-training stage.}
\label{ptr_err}
\end{figure}

%We combine Eq.~\ref{ltrans} and Eq.~\ref{ltar} to 
Combining Eqs.~\ref{ltrans} and~\ref{ltar}, we get the final pre-training loss as follows: % loss: %our loss function for pre-training as follows:
\begin{equation}
\small
    \mathcal{L}_{\text{pre}} =  \mathcal{L}_\text{tar} + \lambda\mathcal{L}_{\text{trans}},
    \label{loss_pre}
\end{equation}
where $\lambda$ is a hyper-parameter to balance two losses. The back-propagation of two losses is shown in Fig.~\ref{ptr_err}.

%The pre-training process uses the simulation data generated by the physics-based model and requires no true observed labels $\textbf{y}$. 
In summary, our pre-training strategy  helps initialize PGRGrN using both simulated intermediate physical variables and simulated target variables. %in two ways. First, %it adds supervision to transferred variables that are passed across segments by leveraging  the intermediate physical variables simulated by PRMS-SNTemp. %Instead of automatically learning a graph convolutional model, our pretraining strategy allows answering 
%This explicitly enforces 
%we 
By enforcing known physical relationships between segments, the model becomes more generalizable.  %by adding supervision on transferred variables. %to be consistent with prior physical knowledge. %what information is propagated among different nodes in the graph. 
%Second, we utilize the simulated target variables $\tilde{\textbf{y}}$ in the loss function $\mathcal{L}_{\text{tar}}$ to pre-train the model. in an end-to-end fashion.  
%The idea is that the 
Besides, the ML model pre-trained with adequate simulated $\tilde{\textbf{y}}$ can get much closer to its final optimal solution, and thus require fewer observation data for fine-tuning. We call this model which is pre-trained using simulation data and then fine-tuned with true observations as PGRGrN$^\text{ptr}$.  

% \yell{Move some of this to the beginning of Section B.}
% % generalize to other applications
% While we discuss the pre-training strategy in the context of river temperature, the proposed pre-training strategy can be applied to a wide range of complex systems with interacting processes. This method can leverage simulated outputs and intermediate variables from physics-based models to pre-train the ML model on its final outputs and transferred variables, respectively. 
% This has a potential to significantly reduce the number of true observations required to fine-tune the model and also enforce the physical relationship between different processes. 

\subsection{Segment-wise contrastive loss}
%However, t
The relationship between input features and target variables can be very complex in environmental systems, e.g., slight changes in %certain features (e.g., 
segment slope and catchment size can drastically alter the streamflow. %the output (e.g., streamflow). %For example, the river segments with small difference in their geometric structure (e.g., width and depth) can have very different streamflow values given similar precipitation. 
Traditional loss functions for regression problems,  such as mean squared loss, tend to be dominated by river segments with larger errors while degrading the performance on other segments with smaller errors. This issue can be further exacerbated given limited observation data on most river segments, especially low-flow segments. Although improving the segments with smaller errors does not contribute much to reducing the overall error, accurately predicting streamflow at these segments provides important information regarding habitat for aquatic life and %aquatic 
biogeochemical cycling.

We introduce a new loss function  to balance the model performance over different segments, with the goal that the global ML model trained on all the river segments should also be consistent with the local patterns extracted from each river segment. In particular, we train $N$ individual LSTM models, $\mathcal{M}_1$ to $\mathcal{M}_N$, for each segment using the simulation data. %These models are trained separately 
Each individual model $\mathcal{M}_i$ is trained to predict simulated target variables (i.e., $\tilde{\textbf{y}}_i$) for a specific segment $i$. % are not impacted by other segments. 
Even though there is a gap between simulation data and true observation data, these individual models have a better chance at capturing the local temporal patterns of each river segment.  %capturing local dynamic patterns that are encoded by the physics-based model. 

When we apply the global PGRGrN model to a specific segment $i$, %we aim to ensure 
the  patterns predicted by the PGRGrN model  should be  similar to the local patterns learned by the individual model $\mathcal{M}_i$. %of this segment. %of the same segment. 
Specifically, we compute the hidden representation $\textbf{h}_{i}^{t}$ from PGRGrN at each time $t$, which encodes dynamic  patterns that directly output target variables. %are directly related to the target outputs. 
Similarly, we run the individual model $\mathcal{M}_i$ to compute its hidden representation $\tilde{\textbf{h}}_{i}^{t}$, which encodes the local temporal patterns  for this segment. %When we apply the PGRGrN to any segment $i$, its 
The hidden representation $\textbf{h}_{i}^{t}$ from PGRGrN should be close to the corresponding local hidden representation $\tilde{\textbf{h}}_{i}^{t}$ and different from the local representation of other segments, i.e., $\tilde{\textbf{h}}_{j}^{t}$ for $j\ne i$. More formally, define a contrastive loss as follows:
\begin{equation}
\small
    \mathcal{L}_{\text{ctr}} = -\frac{1}{NT}\sum_{i}\sum_t \text{log}\frac{ \text{exp} ({\textbf{h}_{i}^{t}}^T \textbf{W}_{ctr} \tilde{\textbf{h}}_{i}^{t}) }{\sum_{j} \text{exp} ({\textbf{h}_{i}^{t}}^T \textbf{W}_{ctr} \tilde{\textbf{h}}_{j}^{t}) },
\end{equation}
where $\textbf{W}_{ctr}$ is model parameters for computing the similarity of hidden representation. To ensure that the hidden representation produced by PGRGrN ($\textbf{h}$) and by individual models ($\tilde{\textbf{h}}$) are comparable, we use shared parameters \{$\textbf{W}_y$, $\textbf{b}_y$\}  in the last layer (Eq.~\ref{prd}) for individual models and the global PGRGrN model and they are fitted when training individual models. 

By combining the contrastive loss and the loss of RGrN (Eq.~\ref{loss_PGRGrN}), we get the fine-tuning loss as follows:
\begin{equation}
\small
    \mathcal{L}_{\text{finetune}} = \mathcal{L}_{\text{RGrN}}+\gamma\mathcal{L}_{\text{ctr}},
    \label{loss_sup}
\end{equation}
where $\gamma$ is a hyper-parameter to balance the supervised loss of PGRGrN and the contrastive loss.

%From another perspective, the 
The proposed contrastive loss provides an alternative way in which different segments are comparable with each other. Traditional loss functions do not perform well for every segment because they are defined on target variables which may vary drastically across different segments. Instead, the contrastive loss matches temporal patterns encoded in the space  of hidden representation, which %greatly 
alleviates this issue. %in the contrastive loss. 

\section{Experimental Results}
\label{sec:exp_res}
We evaluate the proposed method for predicting stream temperature and streamflow %. % streamflow and stream temperature. 
%The proposed method is evaluated
using real-world data collected from the Delaware River Basin, which is an ecologically diverse region and a societally important watershed along the east coast of the United States as it provides drinking water to over 15 million people~\cite{williamson2015summary_backup}. We first describe our dataset and baselines. Then we  discuss the results about the predictive performance using sparse data, the effectiveness of pre-training, the spatial distribution of errors, and model generalization.

\subsection{Dataset and baselines}
\label{sec:dataset}

The dataset is pulled from U.S. Geological Survey's National Water Information System~\cite{us2016national} and  the Water Quality Portal~\cite{read2017water}, the largest standardized water quality data set for inland and coastal waterbodies~\cite{read2017water}. %Streamflow and stream temperature observations are pulled from the these databases and matched to our geospatial modeling fabric. 
Observations at a specific latitude and longitude were matched to river segments that vary in length from 48 to 23,120 meters. The river segments were defined by the national geospatial fabric used for the National Hydrologic Model as described by Regan et al.~\cite{regan2018description_backup}, and the river segments are split up to have roughly a one day water travel time. We match observations to river segments by snapping observations to the nearest stream segment within a tolerance of 250 meters. Observations farther than 5,000 m along the river channel to the outlet of a segment were omitted from our dataset. Segments with multiple observation sites were aggregated to a single mean daily streamflow or water temperature value.    %\yell{Jake: please add a brief description about how this data is collected and relevant data source citations.}

%In particular, w
We study a subset of the Delaware River Basin with 42 river segments that feed into the mainstream Delaware River at Wilmington, DE. We use input features at the daily scale from Oct 01, 1980 to Sep 30, 2016 (13,149 dates). The input features have 10 dimensions which include daily average precipitation, daily average air temperature, date of the year, solar radiation, shade fraction, potential evapotranspiration and  the geometric features of each segment (e.g., elevation, length, slope and width). Air temperature and precipitation values were derived from the Daymet gridded dataset. Other input features (e.g., shade fraction, solar radiation, potential evapotranspiration) are difficult to  measure frequently, and we use values produced by the PRMS-SNTemp model as its internal variables. Water temperature observations were available for 32 segments but the temperature was observed only on certain dates. The number of temperature observations  available 
for each segment ranges from 1 to 9,810 with a total of  51,103 %temperature data observed on different dates and different segments
observations across all dates and segments. Streamflow observations were available for 18 segments. The number of  streamflow observations  available 
for each segment ranges from 4,877 to 13,149 %and in total we have 
with a total of 206,920 observations across all dates and segments. %streamflow data points observed on different dates and different segments. 

% \yell{how to generate adjacency matrix?}

We generate the adjacency matrix $\textbf{A}$ based on the river distance between each pair of river segment outlets, represented as $\text{dist}(i,j)$. We standardize the stream distance  and then compute the affinity level as $\textbf{A}_{ij}=1/(1+\text{exp}(\text{dist}(i,j)))$ for each edge $(i,j)\in\mathcal{E}$.

We compare model performance to multiple baselines, including  the physics-based PRMS-SNTemp model, artificial neural networks (ANN), %recurrent neural networks (RNN) 
RNN with the LSTM cell, and the state-of-the-art %RNN$^\text{ptr}$ 
PGRNN method~\cite{jia2019physics_backup} which uses simulation data to pre-train an LSTM model and then fine-tunes it with true observation data (represented as RNN$^\text{ptr}$). Since a region-specific calibration  PRMS-SNTemp is extremely time-consuming, a version with default values of parameters is widely used in the hydrologic domain~\cite{regan2018description}. We provide a comparison with this version referred to as the PRMS-SNTemp model. % and this is also widely used in the hydrologic domain. %and thus we do not include results of a calibrated model.
We evaluate three variants of the proposed method, RGrN (trained to minimize $\mathcal{L}_\text{RGrN}$), PGRGrN$^\text{ptr}$ (pre-training using the strategy in Section~\ref{sec:ptr} and fine-tuning to minimize $\mathcal{L}_\text{RGrN}$, Eq.~\ref{loss_PGRGrN}), and PGRGrN$^\text{ptr,ctr}$ (pre-training  using the strategy in Section~\ref{sec:ptr} and fine-tuning to minimize $\mathcal{L}_\text{finetune}$, Eq.~\ref{loss_sup}). %In this section, w
%We use RGrN to represent the proposed graph recurrent network directly trained to minimize $\mathcal{L}_\text{RGrN}$ (errors in predictions vs. observation data), and PGRGrN$^\text{ptr}$ for the same model but using the pre-training strategy proposed in Section~\ref{sec:ptr}. Then we use PGRGrN$^\text{ptr,ctr}$ to represent the complete model using both pre-training strategy and the contrastive loss. 
All the ML models are  %using available data over the entire river network 
trained and applied to all the river segments (i.e., all models are global). %The hidden representation in these ML models is in 20 dimension (same for $\textbf{q}_i^t$). 
In the following experiments, we train each ML model using data from the first 24 years (Oct 01, 1980 to Sep 30, 2004) and then test in the next 12 years (Oct 01, 2004 to Sep 31, 2016). The hidden representation in these ML models is in 20 dimension (same for $\textbf{q}_i^t$). We set the learning rate to be 0.0005 and update the model for 200 epochs for modeling water temperature and 300 epochs for modeling streamflow values.

% \noindent\textit{Artificial Neural Networks (ANN)}:

% \noindent\textit{Recurrent Neural Networks (RNN)}:

\subsection{Overall prediction performance}

We report the testing performance of different methods for temperature prediction and streamflow prediction in Table~\ref{perf_temp} and Table~\ref{perf_flow}, respectively. We also test the capacity of each model to learn using less training data %. In particular, we 
by randomly selecting 0.1\% and  2\% %and 10\% training data and then only use the selected 
labeled data from first 24 years for training the model. For RNN$^\text{ptr}$, PGRGrN$^\text{ptr}$ and PGRGrN$^\text{ptr,ctr}$, we 
assume the access to simulation data on every single date from Oct 01 1980 to Sep 20 2016 because they can be generated  by simply running PRMS-SNTemp model on input drivers. %For each experiment, we randomly initialize the model and also randomly select 2\% or 10\% data. 
We repeat each experiment five times with random model initialization and random selection of sparser data (0.1\%, 2\%) and report the mean and standard deviation of the root mean square error (RMSE). %For a better comparison, we also show the performance of uncalibrated PRMS-SNTemp in the first row. Since the calibration of PRMS-SNTemp is extremely time-consuming, the uncalibrated model is widely used in the hydrologic domain and thus we do not include results of a calibrated model.

We can observe that the proposed method outperforms baselines by a considerable margin (Tables~\ref{perf_temp} and~\ref{perf_flow}). The improvement from ANN to RNN shows that the recurrent component is helpful for capturing temporal patterns. RGrN performs better than RNN because it utilizes upstream-downstream dependencies which are critical for an accurate accounting of temperature and streamflow.

We  also observe that  PGRGrN$^\text{ptr}$ has much better performance than RGrN using just 0.1\% or 2\% data. This is because we leverage the physical knowledge to add supervision on multiple model components (i.e., target variables and transferred variables) and thus the model can learn representative latent variables without risking overfitting small amount of observations.   PGRGrN$^\text{ptr,ctr}$ further improves the performance by adjusting the pre-trained model to conform to local patterns of each segment. 
%We also observe that methods using simulation data for pre-training (i.e., RNN$^\text{ptr}$, PGRGrN$^\text{ptr}$ and PGRGrN$^\text{ptr,ctr}$)
The standard pre-training method is also helpful for the RNN model, as we can see the improvement from RNN to RNN$^\text{ptr}$  using simulation data. 

\begin{table}[!t]
\footnotesize
\newcommand{\tabincell}[2]{\begin{tabular}{@{}#1@{}}#2\end{tabular}}
\centering
\caption{Root Mean Squre Error ($\pm$ standard deviation) for temperature modeling using 0.1\%, 2\% and 100\% training labels. Rows in grey color represent methods using simulation data. Here our method is compared with Artificial Neural Networks (ANN), Recurrent Neural Networks (RNN) and the RNN pretrained using simulation data and fine-tuned using observations (RNN$^\text{ptr}$). }
\begin{tabular}{l|cccc}
\hline
\textbf{Method} & 0.1\% & 2\% &  100\% \\ \hline 
PRMS-SNTemp & 3.661&3.661&3.661\\ 
ANN & 3.706$\pm$0.114 &2.159$\pm$0.059 &1.575$\pm$0.035\\
RNN & 3.234$\pm$0.057
& 1.908$\pm$0.048  &1.546$\pm$0.045\\
\rowcolor[gray]{0.95}
RNN$^\text{ptr}$ &2.818$\pm$0.059
&1.810$\pm$0.057  &1.444$\pm$0.039\\ \hline
RGrN &2.849$\pm$0.049&1.906$\pm$0.063 & 1.408$\pm$0.068 \\
\rowcolor[gray]{0.95}
PGRGrN$^\text{ptr}$ &2.556$\pm$0.045&1.715$\pm$0.041  & 1.406$\pm$0.035 \\
\rowcolor[gray]{0.95}
PGRGrN$^\text{ptr,ctr}$& 2.464$\pm$0.105
&1.636$\pm$0.056 & 1.402$\pm$0.034 \\
\hline
\end{tabular}
\label{perf_temp}
\end{table}

\begin{table}[!t]
\footnotesize
\newcommand{\tabincell}[2]{\begin{tabular}{@{}#1@{}}#2\end{tabular}}
\centering
\caption{Prediction RMSE for streamflow modeling using 0.1\%, 2\% and 100\% training labels. Here our method is compared with Artificial Neural Networks (ANN), Recurrent Neural Networks (RNN) and the RNN pretrained using simulation data and fine-tuned using observations (RNN$^\text{ptr}$). }
\begin{tabular}{l|cccc}
\hline
\textbf{Method} &0.1\% & 2\% & 100\% \\ \hline 
PRMS-SNTemp &6.834& 6.834  & 6.834\\ 
ANN &7.116$\pm$0.120&5.777$\pm$0.063 & 4.801$\pm$0.055\\
RNN &6.885$\pm$0.068 &5.718$\pm$0.114 & 4.406$\pm$0.064\\
\rowcolor[gray]{0.95}
RNN$^\text{ptr}$& 6.367$\pm$0.067& 5.529$\pm$0.053 &  4.104$\pm$0.049\\ \hline
RGrN &6.299$\pm$0.053&5.473$\pm$0.064 & 4.139$\pm$0.067 \\
\rowcolor[gray]{0.95}
PGRGrN$^\text{ptr}$ &5.824$\pm$0.075&4.708$\pm$0.032 & 4.106$\pm$0.046 \\
\rowcolor[gray]{0.95}
PGRGrN$^\text{ptr,ctr}$ & 5.895$\pm$0.069&4.679$\pm$0.082 & 4.076$\pm$0.059\\
\hline
\end{tabular}
\label{perf_flow}
\end{table}

\begin{figure} [!t]
\centering
%\raggedleft
\subfigure[]{ \label{fig:b}{}
\includegraphics[width=0.47\columnwidth]{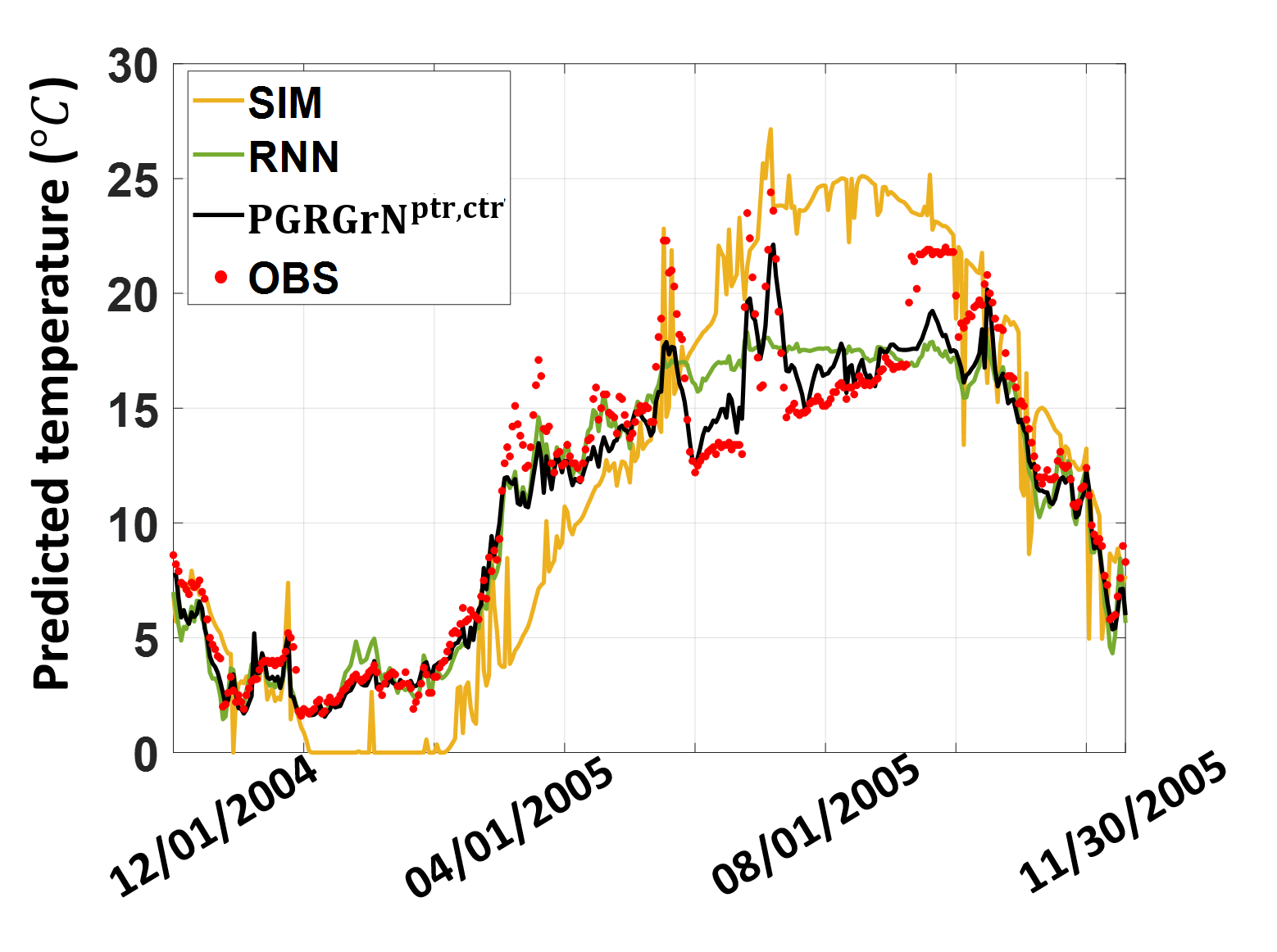}
}\hspace{-0.2in}
\subfigure[]{ \label{fig:b}{}
\includegraphics[width=0.47\columnwidth]{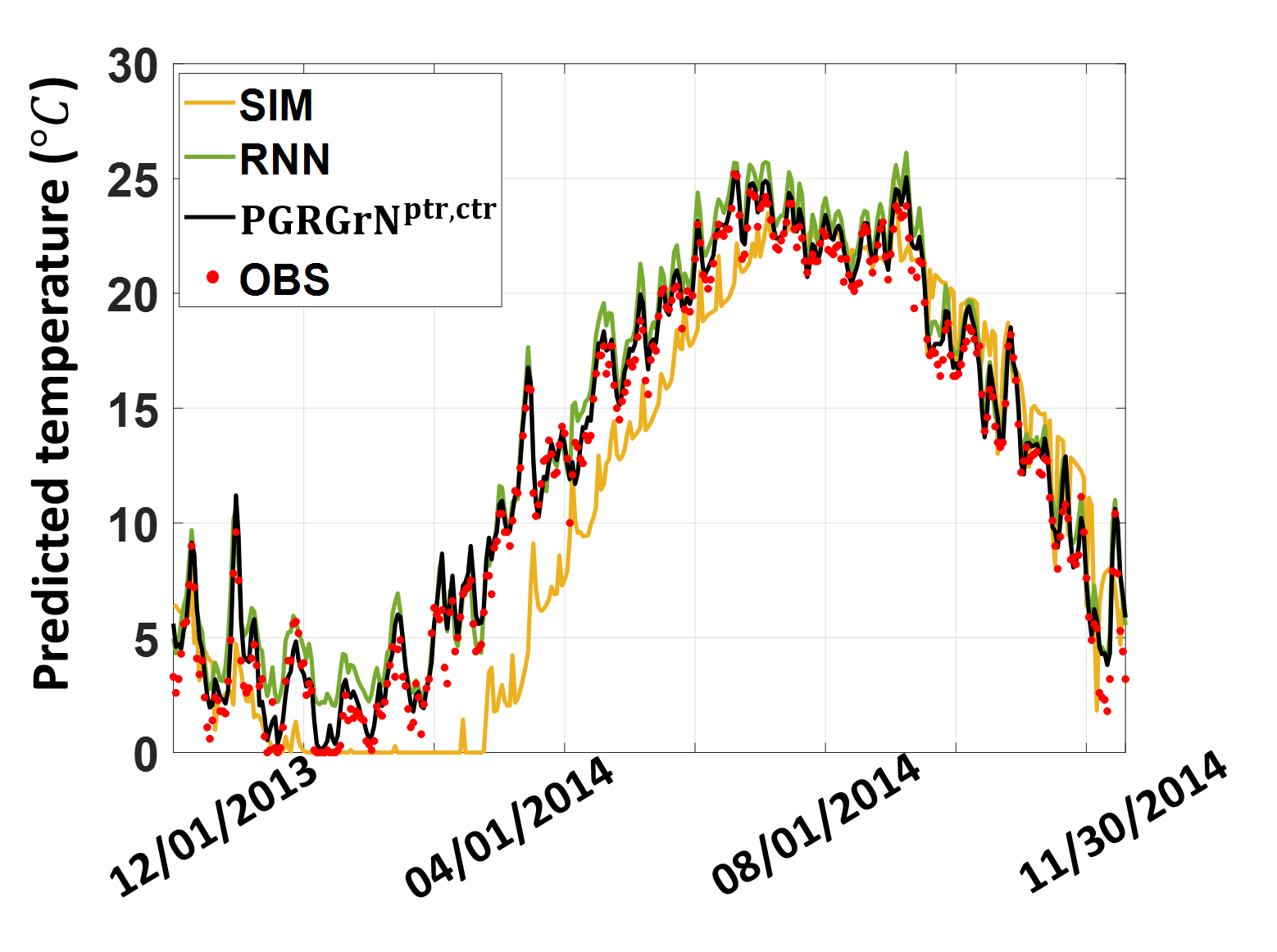}
}\vspace{-.05in}
\subfigure[]{ \label{fig:b}{}
\includegraphics[width=0.47\columnwidth]{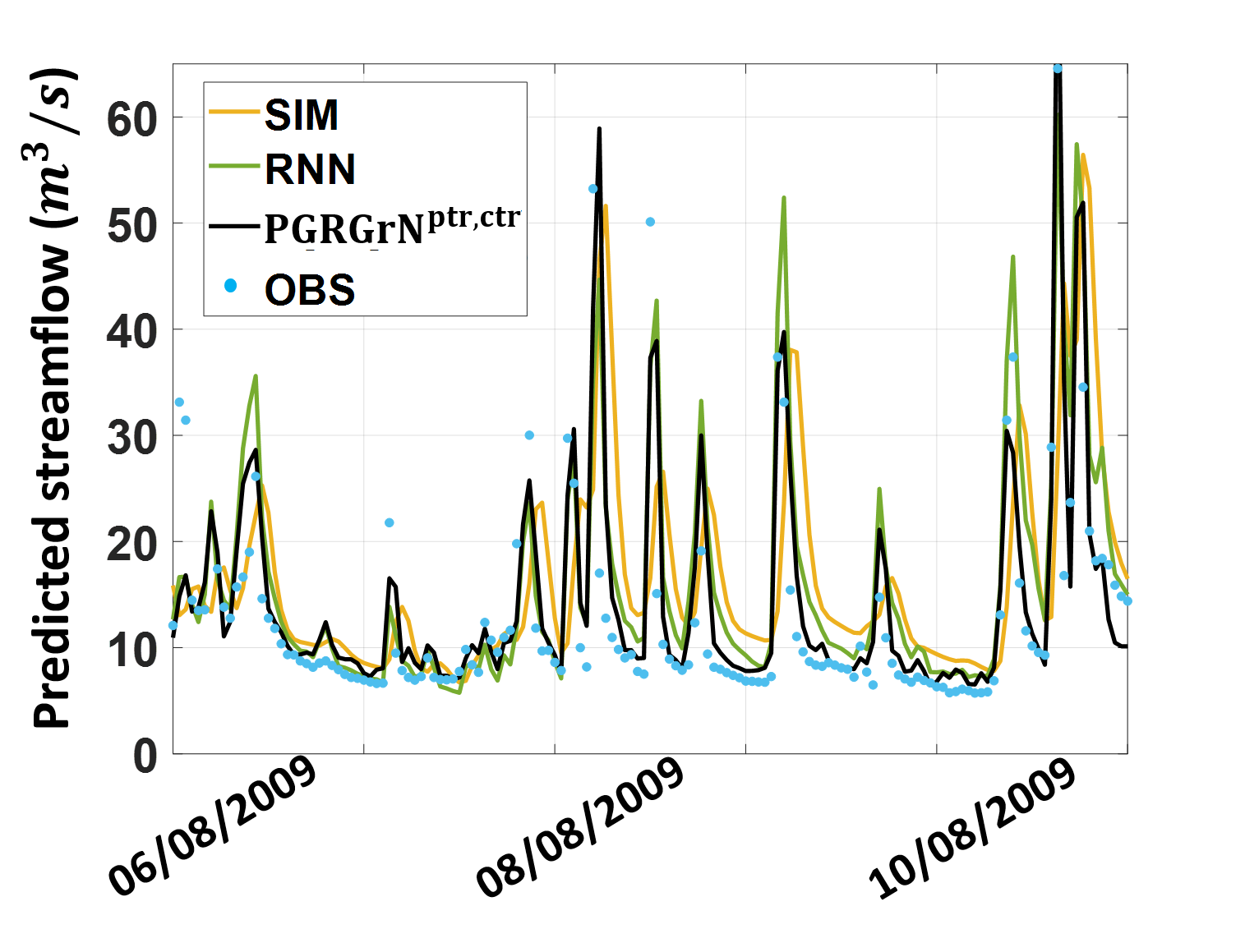}
}\hspace{-0.2in}
\subfigure[]{ \label{fig:b}{}
\includegraphics[width=0.47\columnwidth]{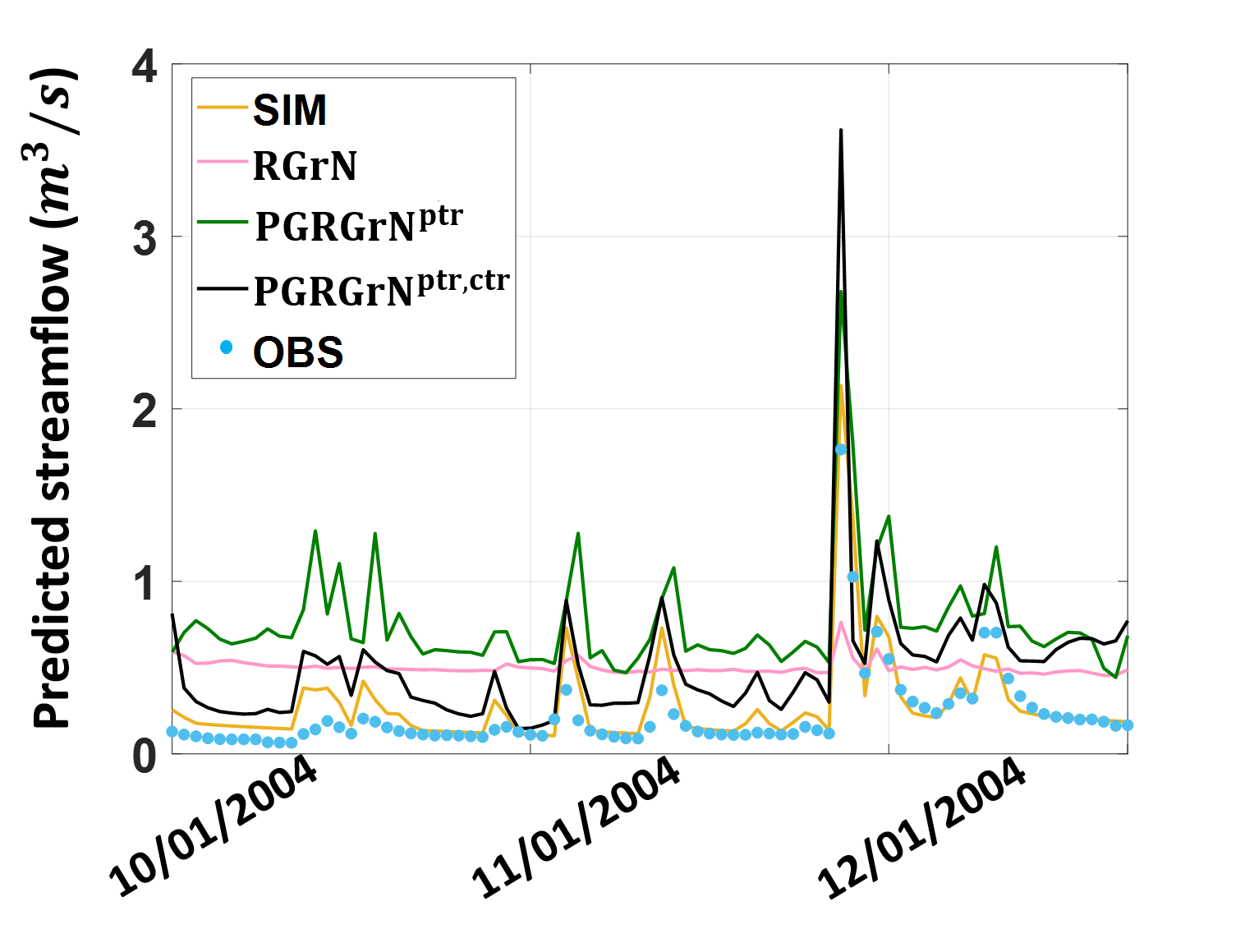}
}\vspace{-.05in}
\caption{(a)(b) Examples of temperature predictions by different methods. (c) An example of streamflow predictions in a high-flow river segment. (d) An example of streamflow predictions by  variants of the proposed method in a low-flow river segment. Here SIM represents the simulated data produced by PRMS-SNTemp model and OBS represents the true observation data. Here we compare the performance of Recurrent Neural Networks (RNN), the proposed model Recurrent Graph Neural Networks (RGrN), the RGrN model using the pretraining strategy (PGRGrN$^\text{ptr}$) and the RGrN model using both pre-training and contrastive loss (PGRGrN$^\text{ptr,ctr}$).  }
\label{exp_pred}
\end{figure}

In Fig.~\ref{exp_pred}, we show several examples for predictions made by PRMS-SNTemp (SIM), RNN, and the proposed model. In temperature prediction (Figs.~\ref{exp_pred} (a) and (b)), PRMS-SNTemp captures the overall dynamic patterns of temperature change but always has   a bias to true observations. RNN gets closer to true observations but does not perform as well as PGRGrN$^
\text{ptr,ctr}$ in capturing fluctuations of temperature changes.

%In streamflow prediction,  W
We can see  similar results for predicting streamflow in segments with high flows (Fig.~\ref{exp_pred} (c)). Here PRMS-SNTemp produces a large bias and also a slow response. %in matching true observations and t
%The proposed method performs 
PGRGrN$^\text{ptr,ctr}$ performs much better than both PRMS-SNTemp and RNN. However, for segments with low streamflows (Fig.~\ref{exp_pred} (d)),  PRMS-SNTemp better matches with observations than ML models. This is because ML models optimize the overall performance while low-flow stream segments (mostly headwaters) are a minority in the entire river networks and contribute less to the loss function. In Fig.~\ref{exp_pred}~(d), we show the results produced by variants of our method to study how the use of contrastive loss and pre-training alter the predictions. We can observe that the predictions made by RGrN are almost constant over time while PGRGrN$^\text{ptr}$ can better capture the dynamics by pre-training the model but has an even larger gap with true observations. By adopting the contrastive loss, PGRGrN$^\text{ptr,ctr}$ effectively reduces the bias on this low-flow segment.

\begin{figure} [!h]
\centering
%\raggedleft
\subfigure[]{ \label{fig:b}{}
\includegraphics[width=0.4\columnwidth]{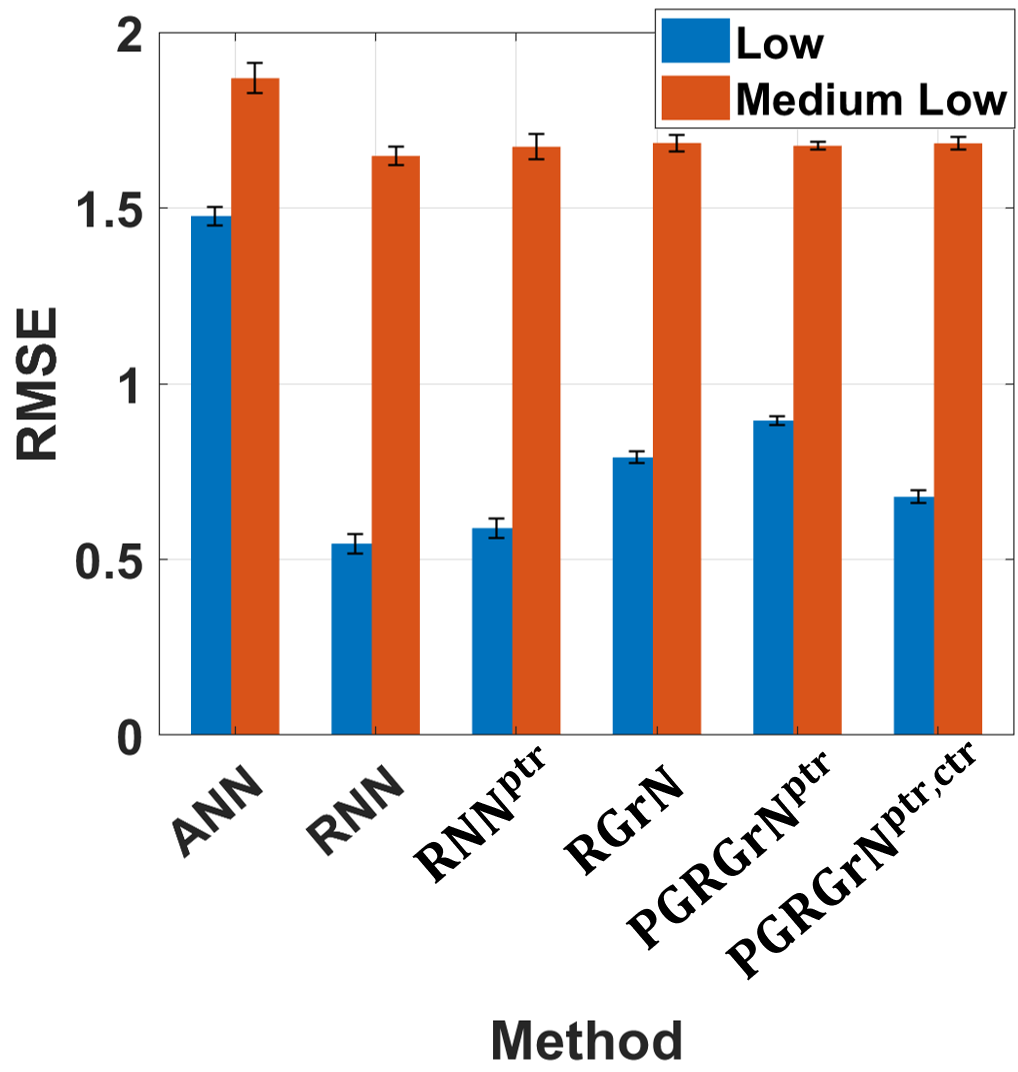}
}\hspace{-0.1in}
\subfigure[]{ \label{fig:b}{}
\includegraphics[width=0.4\columnwidth]{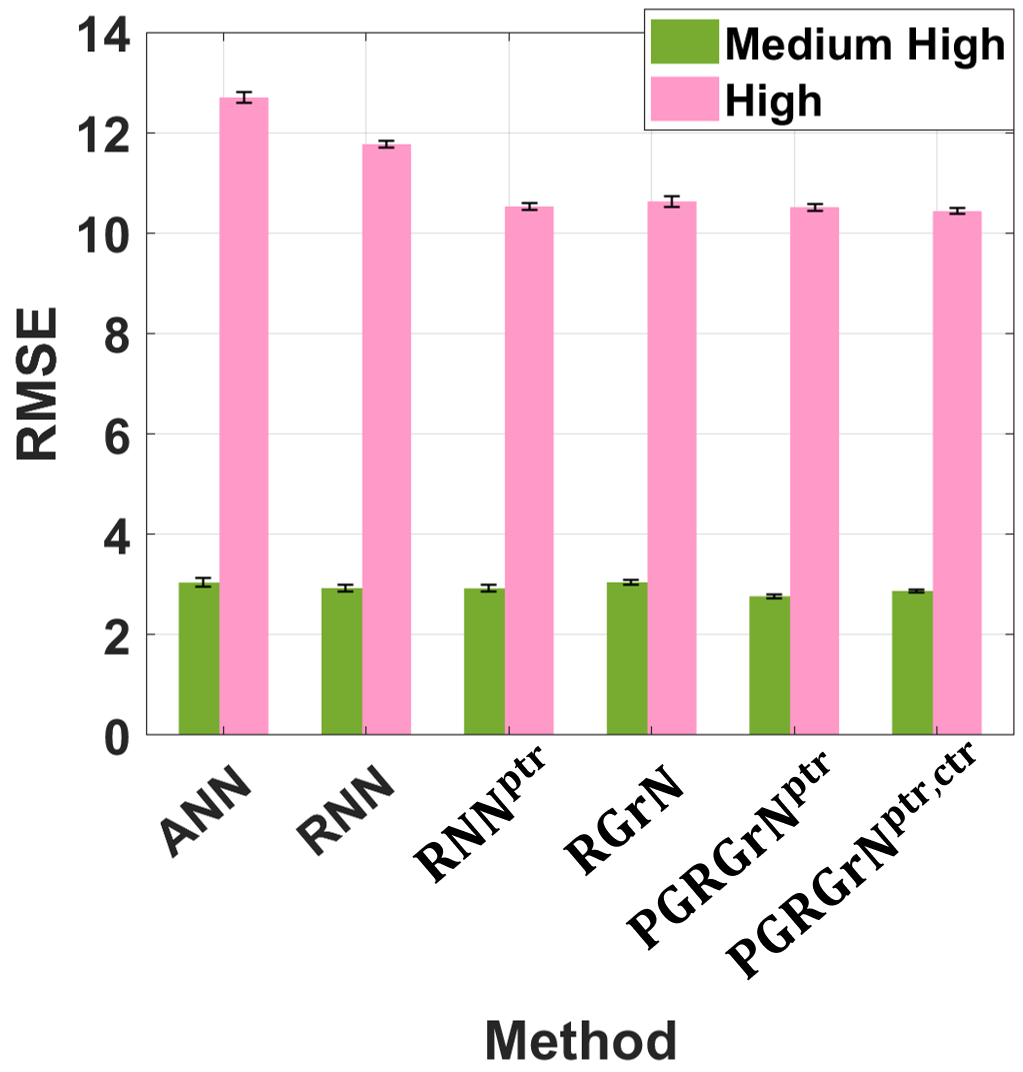}
}\vspace{-.1in}
\caption{Distribution of prediction errors in (a) low and medium-low segments, and (b) medium-high and high-flow segments. }
\label{flow_distr}
\end{figure}
To better understand how the performance varies across different types of river segments, we show the streamflow prediction errors for four types of segments, low ($<$0.5$m^3/s$), medium low (0.5-2$m^3/s$), medium high (2-5$m^3/s$) and high ($>$5$m^3/s$) in Fig.~\ref{flow_distr}. RGrN and PGRGrN$^\text{ptr}$ generally perform better than other methods in medium-high and high-flow river segments but perform much worse in low-flow segments. %because the loss function is dominated by segments with higher flows. 
In particular, we can see RGrN has much larger errors than RNN over low-flow river segments. This is because the neighboring river segments tend to have similar embeddings after graph convolution and thus the training of RGrN pays even less attention to low-flow segments given the fact that there are only a few low-flow segments in the river network. %However, as we can see from 
As shown in Fig.~\ref{flow_distr} (a), this issue is partly addressed by using the contrastive loss.   An alternative solution to this issue is to intelligently select  the most suitable model (e.g., PRMS-SNTemp or ML models) for different types of river segments, which we suggest as  future work.

\subsection{Assessing performance on unobserved segments}
Here we aim to test the performance of models for the segments which have no observation data (Tables~\ref{rm_seg_temp} and~\ref{rm_seg_flow}). Such segments  commonly exist in a real-world basin system. Seg A to Seg E are five river segments which have sufficient observation data for both stream temperature and streamflow. Each row shows the results for an individual experiment where we intentionally remove the temperature or streamflow observations for a specific segment during  the training period (Oct 01, 1980 to Sep 30, 2004). Then we report the prediction performance of RNN, RGrN, and PGRGrN$^\text{ptr,ctr}$ only on this segment during the testing period (Oct 01, 2004 to Sep 31, 2016).

We can observe %that all the models produce 
larger errors produced by all the three models after we remove training data for a 
segment. This is expected because segment-specific observation data has not been used for refinement of the model. However, we observe that the drop in performance of PGRGrN$^\text{ptr,ctr}$ is consistent and often significantly smaller than that of the RNN model.

%This is because the patterns learned from other segments may not be suitable for this target segment. In general, RGrN outperforms RNN after we remove the data because RGrN utilizes the information from the neighborhood to help  capture the patterns of the target segment. The complete version of the proposed method PGRGrN$^\text{ptr,ctr}$ further improves the performance as it leverages the simulation data to guide the learning of river dynamics.

We notice that RGrN may not produce better streamflow predictions than RNN for every segment. For example, Seg A is a head water and has no upstream river segments. Hence, the graph convolution is not helpful for modeling Seg A. Besides, the information propagated from neighbors may not always be helpful if neighboring segments have very different streamflow patterns. A potential area of future research is to develop new graph operators so that each segment can only learn from neighbors that are most relevant. %specific types of neighbors.  

% Our novel ML method outperforms the uncalibrated process-based model across a wide range of data sparsity tests and river segment characteristics. We did not compare to a calibrated PRMS-SNTemp model due to the time consuming nature and often expert tuning needed for precipitation-runoff physics-based models when predicting flow and temperature in a new basin. However, our novel ML model performance compares well to previously calibrated, state-of-the-art physics-based models applied elsewhere. For example, PRMS-SNTemp calibrated to a basin of similar size to the one used in our study produced a mean RMSE in stream temperature of 1.8 C across 20 monitored locations and a coefficient of variation of RMSE (RMSE / mean flow) for streamflow of 1.2 to 2 across monitored locations with comparable mean discharge to our study sites.  This indicates that our novel ML methods are likely performing as well as or better than calibrated physics-based models.   

\begin{table}[!t]
\small
\newcommand{\tabincell}[2]{\begin{tabular}{@{}#1@{}}#2\end{tabular}}
\centering
% this is in the sequence 2,7,8,9,31
\caption{RMSE of temperature prediction on individual segments after removing training observation data. Here we compare the performance of Recurrent Neural Networks (RNN), the proposed Recurrent Graph Neural Networks (RGrN) and the RGrN model using both pre-training and contrastive loss (PGRGrN$^\text{ptr,ctr}$).}
\begin{tabular}{l|l|cccc}
\hline
\textbf{Segment} &  Method & With Obs & Without Obs\\ \hline 
&RNN &2.297$\pm$0.082 & 3.328$\pm$0.132\\ 
Seg A&RGrN& 2.135$\pm$0.060 & 2.749$\pm$0.079\\
&PGRGrN$^\text{ptr,ctr}$&2.084$\pm$0.053 & 2.501$\pm$0.037 \\ \hline
&RNN & 1.116$\pm$0.064 & 1.384$\pm$0.065\\ 
Seg B&RGrN& 0.981$\pm$0.037 & 1.214$\pm$0.032\\
&PGRGrN$^\text{ptr,ctr}$& 1.047$\pm$0.024 & 1.205$\pm$0.016\\ \hline
&RNN & 1.082$\pm$0.083 & 1.804$\pm$0.041\\ 
Seg C&RGrN& 1.013$\pm$0.033 & 1.796$\pm$0.077\\
&PGRGrN$^\text{ptr,ctr}$&0.989$\pm$0.026 & 1.589$\pm$0.040 \\ \hline
&RNN & 0.955$\pm$0.053 & 1.805$\pm$0.064\\ 
Seg D&RGrN& 0.902$\pm$0.026 & 1.597$\pm$0.024\\
&PGRGrN$^\text{ptr,ctr}$&0.996$\pm$0.025 & 1.297$\pm$0.017 \\ \hline
&RNN & 1.067$\pm$0.045 & 1.646$\pm$0.075\\ 
Seg E&RGrN& 0.977$\pm$0.031 & 1.357$\pm$0.033\\
&PGRGrN$^\text{ptr,ctr}$& 1.013$\pm$0.025 & 1.345$\pm$0.041\\ \hline
\end{tabular}
\label{rm_seg_temp}
\end{table}

\begin{table}[!t]
\small
\newcommand{\tabincell}[2]{\begin{tabular}{@{}#1@{}}#2\end{tabular}}
\centering
% this is in the sequence 2,7,8,9,31
\caption{RMSE of streamflow prediction on individual segments after removing training observation data. Here we compare the performance of Recurrent Neural Networks (RNN), the proposed Recurrent Graph Neural Networks (RGrN) and the RGrN model using both pre-training and contrastive loss (PGRGrN$^\text{ptr,ctr}$).}
\begin{tabular}{l|l|cccc}
\hline
\textbf{Segment} &  Method & With Obs & Without Obs\\ \hline 
&RNN & 0.643$\pm$0.074 &0.879$\pm$0.063\\ 
Seg A&RGrN& 0.628$\pm$0.042 &1.271$\pm$0.035\\
&PGRGrN$^\text{ptr,ctr}$& 0.634$\pm$0.053 &0.785$\pm$0.064\\ \hline
&RNN & 3.231$\pm$0.054 & 3.493$\pm$0.112\\
Seg B&RGrN& 3.185$\pm$0.066&3.234$\pm$0.045\\
&PGRGrN$^\text{ptr,ctr}$& 2.981$\pm$0.067&3.043$\pm$0.091\\ \hline
&RNN & 1.798$\pm$0.075 &2.151$\pm$0.094 \\
Seg C&RGrN& 1.749$\pm$0.071&2.139$\pm$0.085\\
&PGRGrN$^\text{ptr,ctr}$& 1.697$\pm$0.065&1.895$\pm$0.061\\ \hline
&RNN & 1.983$\pm$0.103 & 2.711$\pm$0.114\\ 
Seg D&RGrN& 2.006$\pm$0.110&2.354$\pm$0.077\\
&PGRGrN$^\text{ptr,ctr}$&1.989$\pm$0.084&2.343$\pm$0.079\\ \hline
&RNN & 10.618$\pm$0.082 &11.828$\pm$0.073\\ 
Seg E&RGrN& 10.833$\pm$0.070&12.078$\pm$0.075\\
&PGRGrN$^\text{ptr,ctr}$& 9.726$\pm$0.052& 10.803$\pm$0.103\\ \hline
\end{tabular}
\label{rm_seg_flow}
\end{table}

\begin{figure} [!h]
\centering
%\raggedleft
%\subfigure[]{ \label{fig:b}{}
%\includegraphics[width=0.47\columnwidth]{abl_temp_seg7_2009.png}
%}\hspace{-0.2in}
\subfigure[]{ \label{fig:b}{}
\includegraphics[width=0.47\columnwidth]{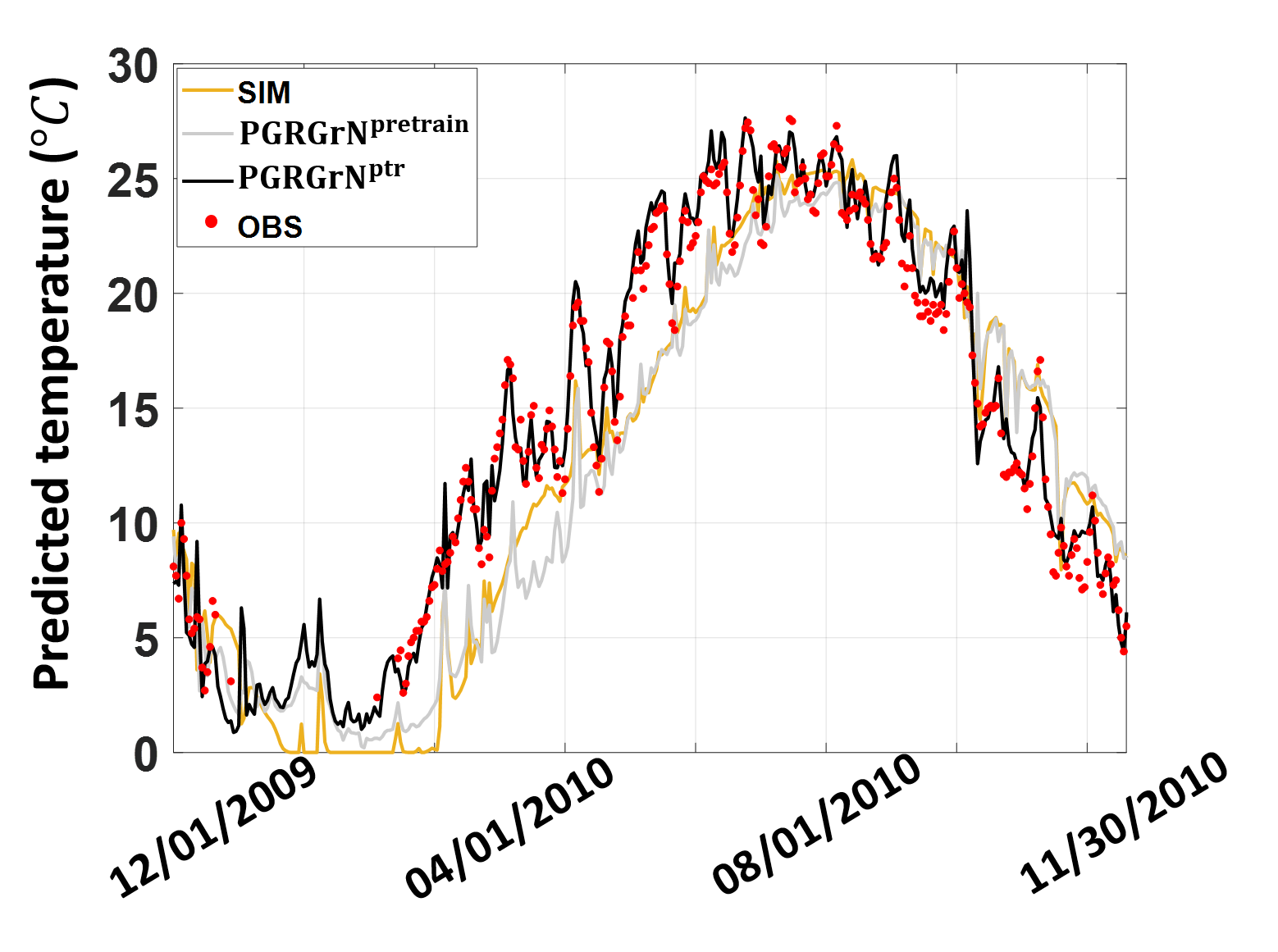}
}\hspace{-0.2in}
%\subfigure[]{ \label{fig:b}{}
%\includegraphics[width=0.47\columnwidth]{abl_flow_seg2.png}
%}\hspace{-0.2in}
\subfigure[]{ \label{fig:b}{}
\includegraphics[width=0.47\columnwidth]{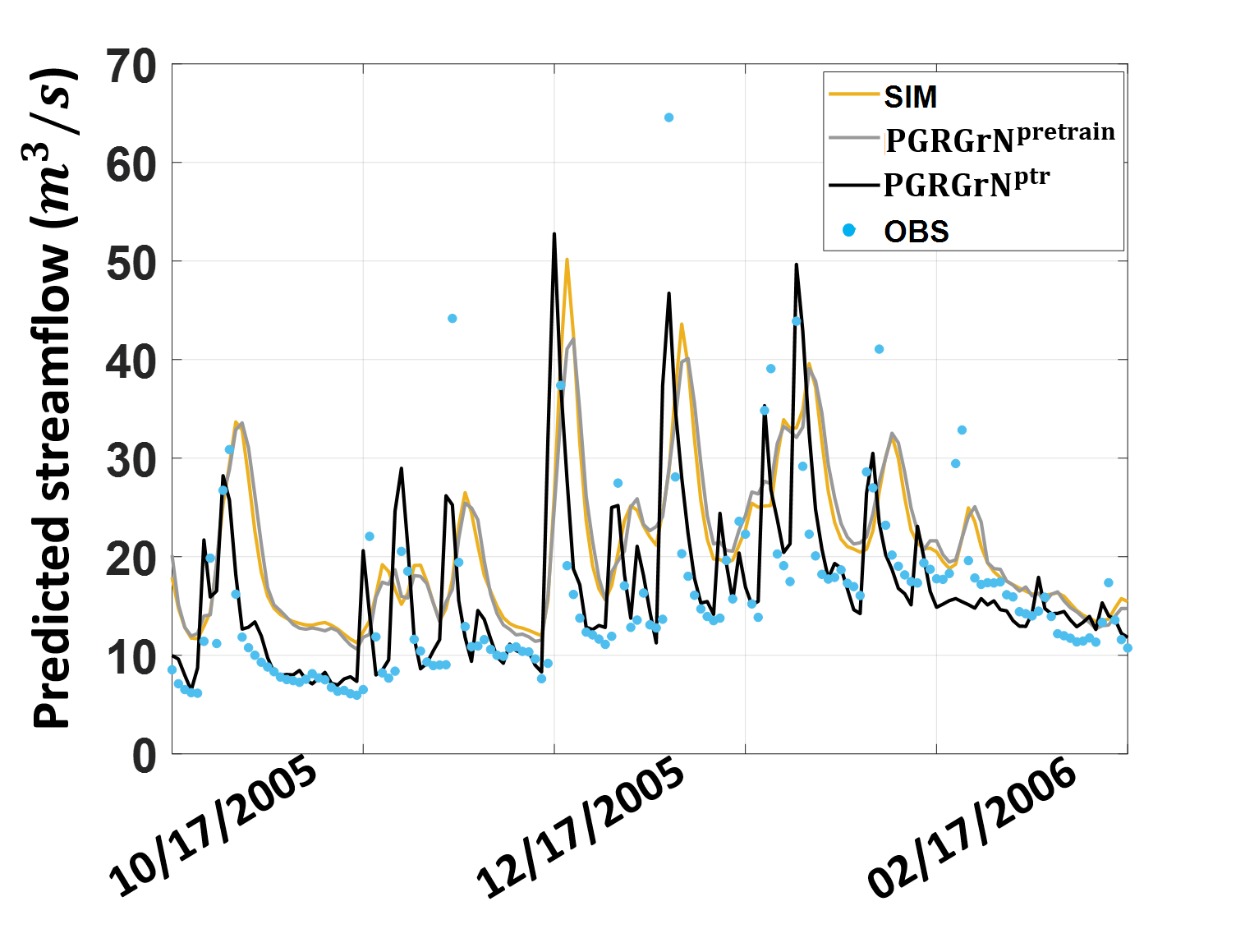}
}\vspace{-.1in}
\caption{The predictions of pre-trained (PGRGrN$^\text{pretrain}$) and fine-tuned models (PGRGrN$^\text{ptr}$)  for (a) temperature and (b) streamflow, using 2\% training data.  Here we compare the performance of the RGrN model that is only pre-trained using simulation data (PGRGrN$^\text{pretrain}$) and the RGrN model using pre-training and then fine-tuned with observation data (PGRGrN$^\text{ptr}$). }
\label{exp_ptr}
\end{figure}

\subsection{Effectiveness of pre-training}
%Now we illustrate how pre-training guide the learning process. 
In Fig.~\ref{exp_ptr}, we randomly select two %segments as examples 
example segments to show how predictions change from the pre-trained model (PGRGrN$^\text{pretrain}$) to the fine-tuned model (PGRGrN$^\text{ptr}$)  using 2\% training data. %For temperature predictions (Fig.~\ref{exp_ptr} (a)), 
Here the model PGRGrN$^\text{ptr}$ is the same with the PGRGrN$^\text{ptr}$ model used in the previous results but is only fine-tuned using 2\% observations. In contrast, PGRGrN$^\text{pretrain}$ is only pre-trained using the strategy proposed in Section~\ref{sec:ptr} but without using observations for fine-tuning. 
Fig.~\ref{exp_ptr} (a) %, we can see 
shows that the pre-trained model  match the simulated temperatures very well and thus can capture  general temperature patterns even without using observation data. There is still a gap between true observations and %the predictions made by the pre-trained model 
PGRGrN$^\text{pretrain}$ since PGRGrN$^\text{pretrain}$  emulates PRMS-SNTemp, % simulation data created by physics-based model have 
which has inherent bias due to an incomplete representation of physics. %However, since the pre-trained model already captures the general patterns, 
Nevertheless, after learning general patterns from simulation data, %it 
the model can be fine-tuned to match true observations using %requires 
much less training data. %to fine-tune it to match true observations. 
This can be verified as we show that PGRGrN$^\text{ptr}$ fine-tuned with 2\% data closes the gap between PGRGrN$^\text{pretrain}$ and true observations. Similarly, %In streamflow prediction, we can observe similar results in
Fig.~\ref{exp_ptr} (b) shows that pre-training helps %initialize the model to 
capture general streamflow patterns and  fine-tuning effectively fixes the bias and the slow response of PRMS-SNTemp. %the physics-based model.% made by the simulation data. %In streamflow prediction, we can observe similar results in Fig.~\ref{exp_ptr} (b) where the fine-tuning effectively fixes the gap and the slow response made by the simulation data. %However, in low-flow segments (Fig.~\ref{exp_ptr} (c))  we can observe that the pre-trained and fine-tuned model does not perform as well as the simulation data because their training loss pays less attention to such segments. This also motivates the use of contrastive loss.

%\yell{(c) motivates the use of contrastive loss.}

\begin{figure*} [!h]
\centering
%\raggedleft
\subfigure[]{ \label{fig:b}{}
\includegraphics[width=0.95\columnwidth]{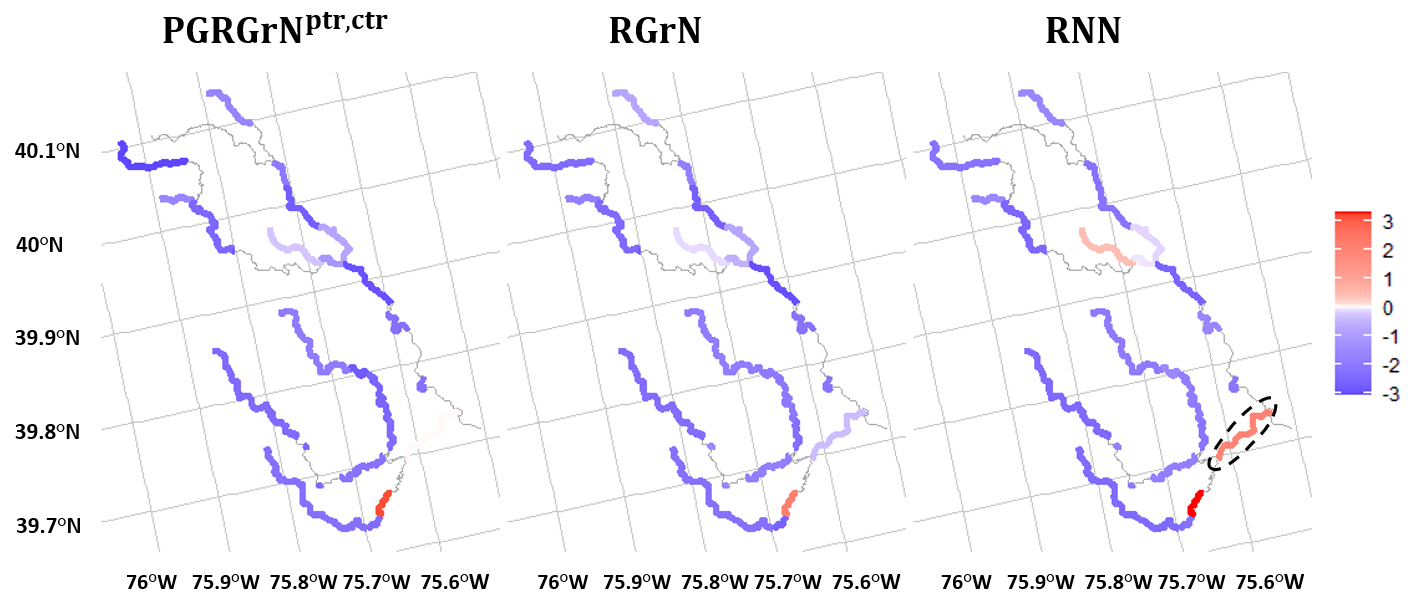}
}%\hspace{-0.2in}
\subfigure[]{ \label{fig:b}{}
\includegraphics[width=0.95\columnwidth]{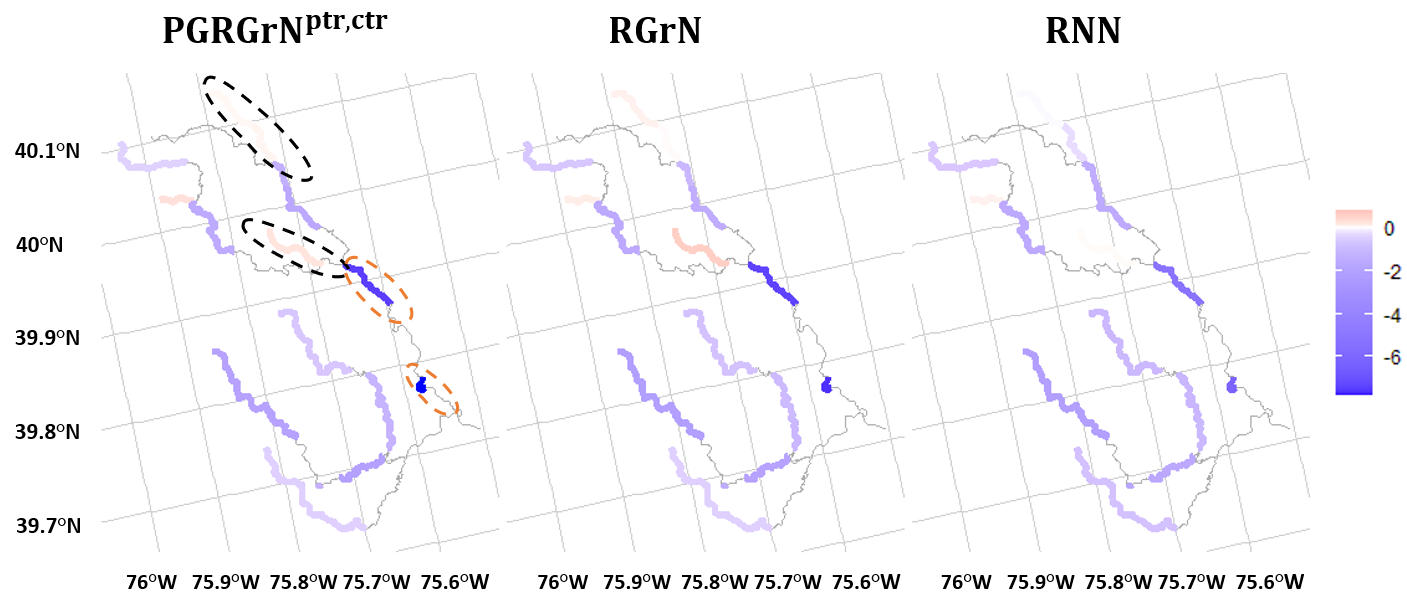}
}\vspace{-.05in}
\caption{Prediction errors across different segments for (a) temperature prediction and (b) streamflow prediction. Here we show the error as the RMSE of each method minus the RMSE of PRMS-SNTemp simulation data to get a better contrast. Darker blue indicates better performance of the ML model over the physics-based model. In (a), we do not discuss the red colored segment on the bottom (predicted poorly by all the three methods) and  head water (75.7$^\circ$W, 39.95$^\circ$N) predicted poorly by RNN because they just have one test observation. }
\label{sp}
\end{figure*}

\subsection{Spatial distribution of errors}
In Fig.~\ref{sp}, we show the distribution  across different segments for ML temperature and streamflow prediction improvements over the physics-based PRMS-SNTemp simulations. In Fig.~\ref{sp}~(a), we can observe that RGrN and PGRGrN$^\text{ptr,ctr}$ produce smaller temperature error than RNN in many segments. We find one segment (in dashed circle) where RNN produces worse predictions than PRMS-SNTemp but RGrN greatly improves the prediction. This is the only segment in the data set for which we have no training data but sufficient testing data. The reason why RGrN can produce better predictions for this unlabeled segment is that it leverages the dependencies with other segments to learn the temperature patterns even without training data from this specific segment.

For streamflow prediction (Fig.~\ref{sp} (b)), it can be seen that RGrN and PGRGrN$^\text{ptr,ctr}$ have lower RMSE in several high-flow segments (e.g., the segments in red dashed circles). However, RGrN performs worse than RNN on  headwater segments (in black dashed circles) and these segments have lower streamflow. Compared with RGrN, it can be seen that PGRGrN$^\text{ptr,ctr}$ alleviates this issue and produces smaller errors in these low-flow segments  by using the contrastive loss.

\subsection{Generalization test}
%\yell{Change the text here to match with the updated tables.}
%In scientific applications, it is critical to test model generalizability  
Generalizability is important for scientific problems because most observation data can be collected from certain periods or locations for which it is easier to deploy sensors. Hence, we test model generazability for both streamflow and stream temperature prediction. In particular, we  test the model performance on streamflow prediction %when we have 
using training data only from segments with higher average streamflows ($>$3$m^3/s$) in the first 24 years (Oct 1980 to Sep 2004). Then we report the  RMSE in %testing years 
in the next 12 years for segments with lower average streamflows ($<$1$m^3/s$) 
in Table~\ref{gen_flow}. We also include the testing performance of the model trained using all the observations from first 24 years as a baseline in the second column of Table.~\ref{gen_flow}. Similarly, %it is  known that 
%temperature varies drastically across different seasons and 
it is challenging for traditional ML models to generalize to  predicting temperatures in a season that was not included in training data. %seasons other than the training seasons. 
We train each model using data only from colder seasons (spring, fall and winter) in the first 24 years and then test in summers in the next 12 years, as shown in Table~\ref{gen_temp}.
%, we separately report the testing performance in summer and other seasons in the testing years.

Note that PGRGrN$^\text{ptr,ctr}$ always performs better than other methods because the incorporation of physical knowledge and minimizing the  contrastive loss forces the model to learn generalizable patterns for each segment. %get closer to local patterns of each segment.
It can be seen that ANN has large errors (especially in Table~\ref{gen_flow}) when applied to scenarios that are very different to training data. This is because ANN only focuses on the mapping from input features to target variables while physical phenomena are often driven by  spatial and temporal processes. Such a gap makes ANN prone to overfit the training data. In contrast, RGrN performs better than ANN and RNN in temperature prediction because it leverages the dependencies both over space and  time, and thus has a better chance in learning generalizable  patterns. However, the performance of RGrN becomes worse in predicting segments with lower streamflows. % for low-flow segments. %The graph modeling tends to smooth out the 
%The 
%because t
The graph convolution makes the extracted hidden representation ($\textbf{h}_i^t$) %over space and neighboring segments tend to have 
more similar for neighboring segments and thus  %hidden representation  and thus 
RGrN  simulate streamflows with similar dynamic patterns as in high-flow segments when it is trained just using observation data from high-flow segments.   

%The models that use simulation data generally perform much better in generalization tests because they utilize the simulation data on every single date to learn generalizable patterns. Moreover, PGRGrN$^\text{ptr,ctr}$ always performs better than other methods because minimizing the  contrastive loss forces the model to get closer to local patterns of each segment.

\begin{table}[!h]
\small
\newcommand{\tabincell}[2]{\begin{tabular}{@{}#1@{}}#2\end{tabular}}
\centering
\caption{Streamflow RMSE on segments with streamflow $<$1$m^3/s$ from 2005 to 2016. Each model is trained using observation data from segments with streamflow $>$3$m^3/s$ (column 1) or all the observations data (column 2) from Oct 1980 to Sep 2004. Here our method is compared against Artificial Neural Networks (ANN),   Recurrent Neural Networks (RNN), the RNN that is pre-trained using simulation data and then fine-tuned using observation data (RNN$^\text{ptr}$).}
\begin{tabular}{l|cccc}
\hline
\textbf{Method} & Train on high-flow & Train on all the data \\ \hline 
%SNTemp & 1.419&11.181\\ 
ANN &9.752$\pm$0.527 & 1.719$\pm$0.114\\
RNN &3.012$\pm$0.432 & 1.196$\pm$0.054\\
\rowcolor[gray]{0.95}
RNN$^\text{ptr}$ & 3.279$\pm$0.526 & 1.075$\pm$0.063\\ \hline
RGrN & 4.073$\pm$0.082 & 1.212$\pm$0.066  \\
\rowcolor[gray]{0.95}
PGRGrN$^\text{ptr}$ & 2.226$\pm$0.089 & 1.206$\pm$0.076  \\
\rowcolor[gray]{0.95}
PGRGrN$^\text{ptr,ctr}$ & 2.099$\pm$0.088 & 1.201$\pm$0.089\\
\hline
\end{tabular}
\label{gen_flow}
\end{table}
\begin{table}[!h]
\small
\newcommand{\tabincell}[2]{\begin{tabular}{@{}#1@{}}#2\end{tabular}}
\centering
\caption{Temperature RMSE in summers from 2005 to 2016. Each model is trained using observation data from colder seasons (Column 1) or all the observations data (column 2) from Oct 1980 to Sep 2004. Here our method is compared against Artificial Neural Networks (ANN),   Recurrent Neural Networks (RNN), the RNN that is pre-trained using simulation data and then fine-tuned using observation data (RNN$^\text{ptr}$).%Rows in grey color represent methods using simulation data. 
}
\begin{tabular}{l|cccc}
\hline
\textbf{Method} & Train on cold seasons & Train on all the data \\ \hline 
%SNTemp & 2.944 & 3.906 \\ 
ANN & 2.138$\pm$0.093 & 1.794$\pm$0.032\\
RNN & 2.104$\pm$0.080 & 1.789$\pm$0.034\\
\rowcolor[gray]{0.95}
RNN$^\text{ptr}$ & 1.893$\pm$0.085 & 1.555$\pm$0.021\\ \hline
RGrN & 1.939$\pm$0.062 & 1.539$\pm$0.024\\
\rowcolor[gray]{0.95}
PGRGrN$^\text{ptr}$ &   1.853$\pm$0.034 & 1.530$\pm$0.014\\
\rowcolor[gray]{0.95}
PGRGrN$^\text{ptr,ctr}$ &1.744$\pm$0.053& 1.416$\pm$0.019 \\
\hline
\end{tabular}
\label{gen_temp}
\end{table}

\subsection{Sensitivity tests}
\label{sec:sens}
Here we test the sensitivity of the model to different hyper-parameter settings. Also we test the performance of the model with different time delays (Eq.~\ref{conv}).

\begin{figure} [!h]
\centering
\subfigure[]{ \label{fig:b}{}
\includegraphics[width=0.4\columnwidth]{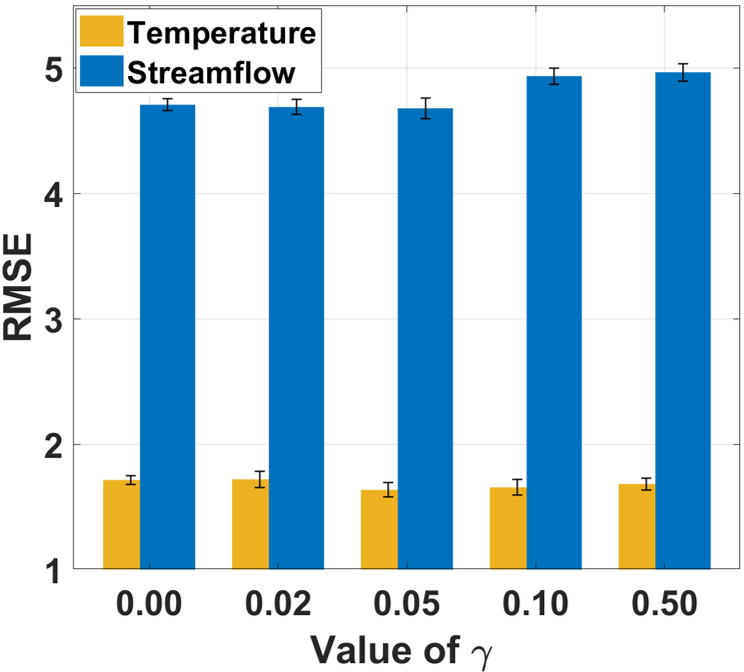}
}\hspace{-0.1in}
\subfigure[]{ \label{fig:b}{}
\includegraphics[width=0.4\columnwidth]{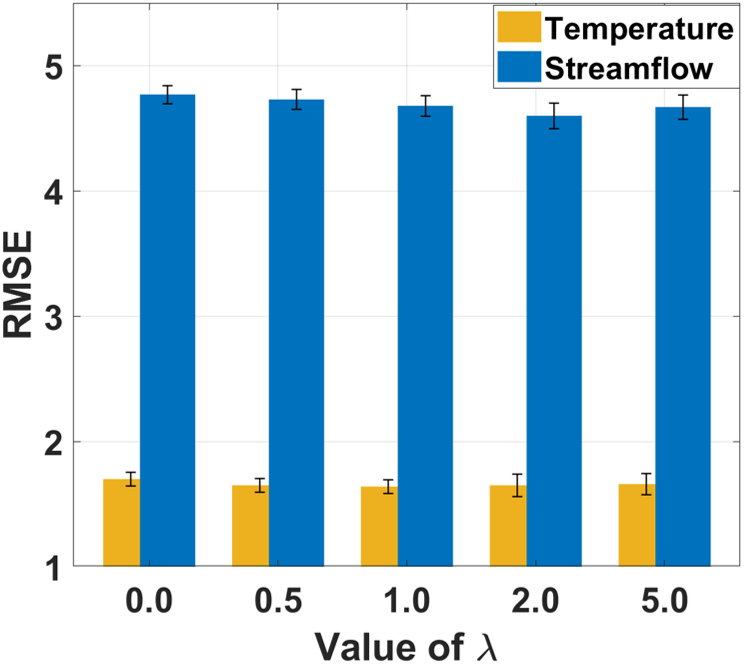}
}\vspace{-.1in}\\
\subfigure[]{ \label{fig:b}{}
\includegraphics[width=0.4\columnwidth]{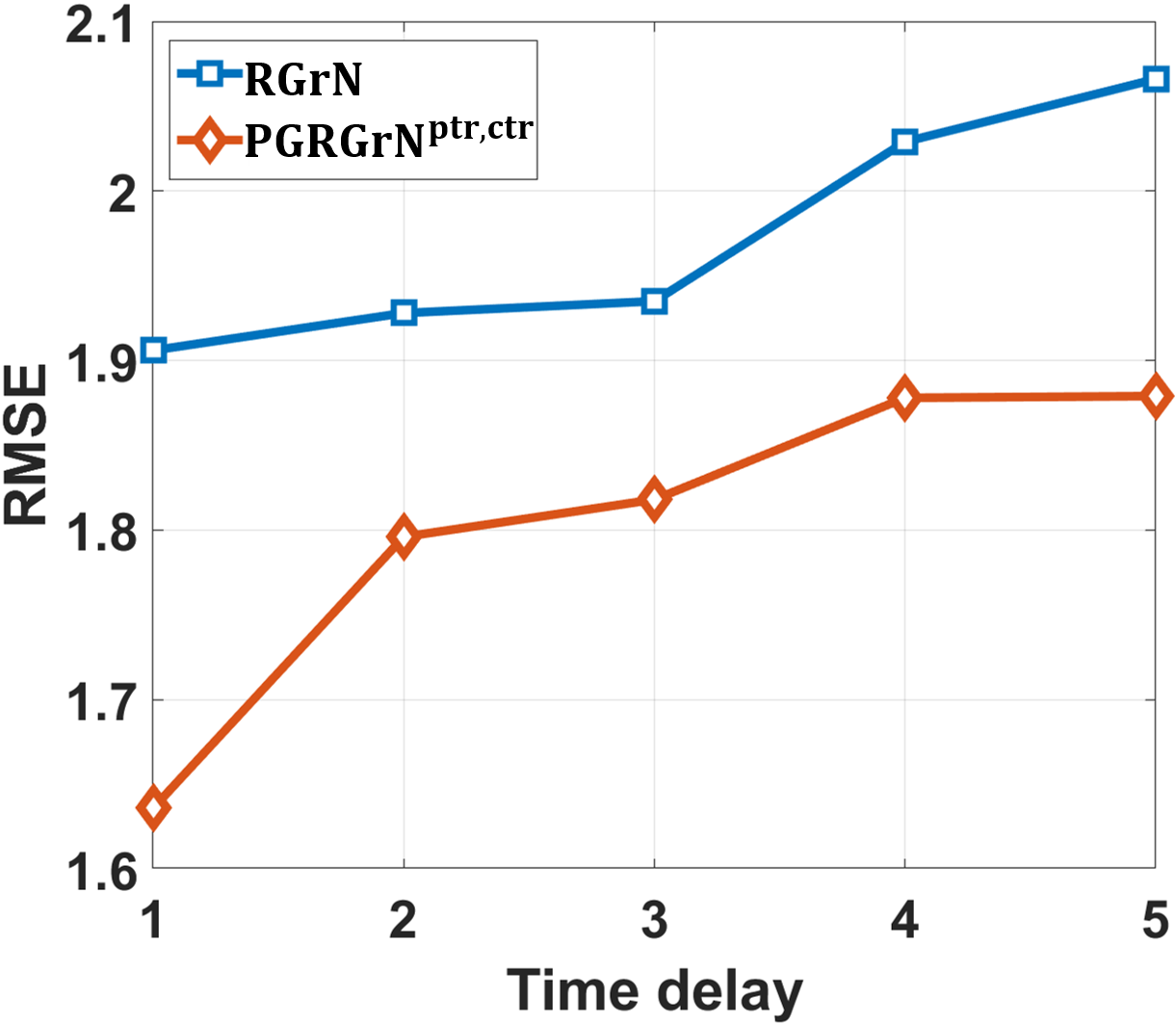}
}\hspace{-0.1in}
\subfigure[]{ \label{fig:b}{}
\includegraphics[width=0.4\columnwidth]{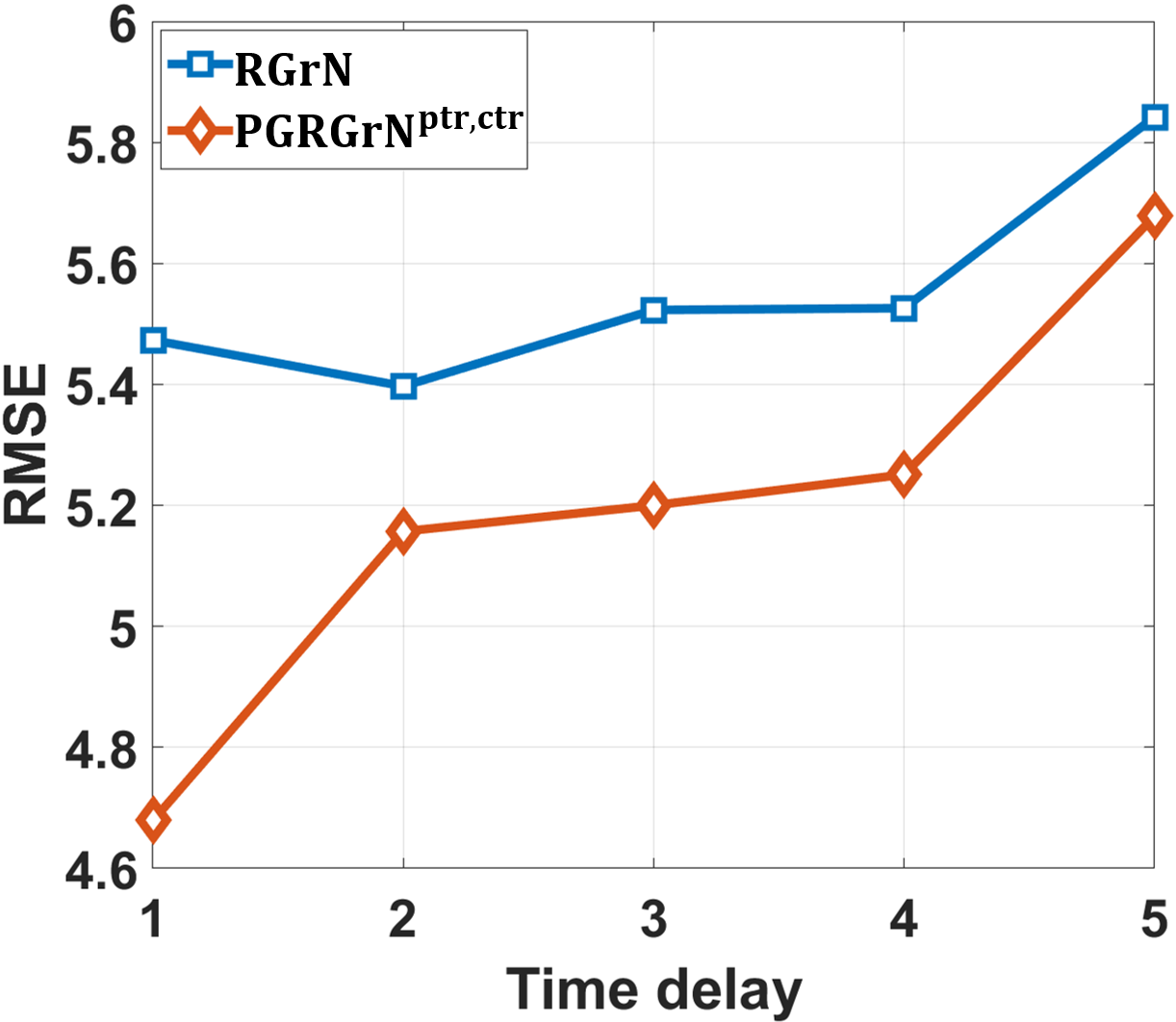}
}\vspace{-0.1in}
\caption{Performance using different values of (a) $\gamma$ and (b) $\lambda$. Performance  with different time delay for (c) temperature and (d) streamflow prediction. In (c) and (d) we sohw the performance of the proposed Recurrent Graph Neural Networks (RGrN) and the RGrN model using both pre-training and contrastive loss (PGRGrN$^\text{ptr,ctr}$).}
\label{sens}
\end{figure}

In Figs.~\ref{sens} (a) and (b), we show the prediction performance using 2\% data using different values of $\lambda$ (Eq.~\ref{loss_pre}) and $\gamma$ (Eq.~\ref{loss_sup}). We can see that the model is relatively robust with different values of hyper-parameters but there is a slight increase of RMSE if we set $\gamma$ and $\lambda$ to be either too small or too large. With a small $\gamma$, the model has less weight on contrastive loss and thus produces worse performance using limited (2\%) observation data. With a large $\gamma$ value, we put a higher weight on the contrastive loss but the model is more likely to be impacted by the bias in the simulation data. On the other hand, the smaller value of $\lambda$ will make the model ignorant of the  supervision on transferred variables while a much larger value of $\lambda$ results in less focus on the target variables. %and parameters in the prediction layer ($\textbf{W}_y$ and $\textbf{b}_y$). 
It is worth mentioning that our adopted hyper-parameters may not be optimal values in the testing set because they are tuned using the simulation data.  %These hyper-parameters $\lambda$ and $\gamma$ are tuned using simulation data

In Figs.~\ref{sens} (c) and (d), we use different lengths of time delay, from one time step to five time steps, in Eq.~\ref{conv}. The time delay indicates how many days prior the information from upstream segments influence the downstream segments. We can see a general increase of RMSE when we set a larger time delay.

\section{Conclusion}
In this paper, we propose a novel  method PGRGrN for modeling interacting segments in river networks. We leverage the prior physical knowledge about segment-to-segment interactions embedded in physics-based models  to enhance the learning of latent representation in the proposed ML model. Moreover, we improve the loss function to optimize both the overall performance over the river network and the local performance on each individual river segment. We have demonstrated the superiority of the proposed method in handling the scarcity of labeled data and in generalizing to unseen scenarios. %We have also studied the spatial distribution of %predictions 
%errors and %illustrated the 
%the effectiveness of pre-training. %strategy. 

In addition to modeling variables in river networks, the proposed method can be adjusted to model other complex systems which involve interacting processes. For example, this method could be potentially used for material discovery, biological research and quantum chemistry to capture interactions between different atoms or molecules. 

While our method performs much better than existing models, it remains limited in precisely predicting special segments (e.g., unobserved segments and segments with extremely low streamflows). To advance understanding, future ML modeling efforts may consider uncertainty of the global ML model and 
determine whether ML should be used to replace physics-based models in different situations.  

\section{Acknowledgments}
Any use of trade, firm, or product names is for descriptive purposes only and does not imply endorsement by the U.S. Government. 
%\yell{Limitations and future works: how to intelligently select different type of method for different components.}

\bibliographystyle{IEEEtran}
\bibliography{IEEEabrv}

\end{document}